\documentclass[onecolumn,showpacs,amsmath,amssymb]{revtex4}
\usepackage{graphicx}
\usepackage{subfigure}
\usepackage{bm} 

\begin{document}
 
\title{On the formation of Hubble flow in Little Bangs}

\author{Miko\l aj Chojnacki} \email{chojnacki_mikolaj@tlen.pl}
\affiliation{Institute of
  Physics, \'Swi\c{e}tokrzyska Academy, PL-25406 Kielce, Poland}
\author{Wojciech Florkowski}
\email{florkows@amun.ifj.edu.pl} 
\affiliation{Institute of
  Physics, \'Swi\c{e}tokrzyska Academy, PL-25406 Kielce, Poland}
\affiliation{Institute of Nuclear Physics, Polish Academy of Sciences, 
  PL-31342 Krak\'ow, Poland} 
\author{Tam\'as Cs\"org\H{o}}
\email{csorgo@sunserv.kfki.hu} \affiliation{MTA KFKI RMKI, H-1525 Budapest 114, P.~O.~Box 49, 
Hungary}

\vspace{1cm}

\date{October 7, 2004}

\begin{abstract}
A dynamical appearance of scaling solutions in the relativistic hydrodynamics applied 
to describe ultra-relativistic heavy-ion collisions is studied. We consider the 
boost-invariant cylindrically symmetric systems and the effects of the phase transition 
are taken into account by using a temperature dependent sound velocity inferred from the 
lattice simulations of QCD. We find that the transverse flow acquires the scaling form 
$r/t$ within the short evolution times, 10-15 fm, only if the initial transverse 
flow originating from the pre-equilibrium collective behavior is present at the 
initial stage of the hydrodynamic evolution. The amount of such pre-equilibrium 
flow is correlated with the initial pressure gradient; larger initial gradients 
require smaller initial flow. The results of the numerical calculations support the 
phenomenological parameterizations used in the Blast-Wave, Buda-Lund, and Cracow models 
of the  freeze-out process.
\end{abstract}



\maketitle 

\section{Introduction \label{intro}}

Hadronic data collected in the heavy-ion experiments at RHIC support the idea
that the system formed in Au+Au collisions is highly thermalized and undergoes
strong transverse and longitudinal expansion \cite{QM02}. Moreover, successful 
parameterizations of the freeze-out process indicate that such expansion may be 
characterized by the Hubble law \cite{Broniowski:2001we,Broniowski:2001uk,Csanad:2004mm}. 
In the simplest form, used in cosmology, this law states the proportionality of the
relative velocity of galaxies to their relative separation,
\begin{equation}
{\bf v} = H {\bf r} .
\end{equation}
The constant of proportionality or Hubble's constant H is a function of time.
In the Friedmann universe~\cite{allday} as well as in analytic solutions of 
non-relativistic fireball hydrodynamics~\cite{zbg}, its value is
\begin{equation}
 H =  \frac{\dot R}{R},
\end{equation}
where the function $R(t)$ is in general a complicated function of time.
In nuclear hydrodynamics, the Hubble law characterizes the fluid velocity
distributions of the expanding matter, while in cosmology the Hubble
law characterizes the expansion of the space.
In the non-relativistic hydrodynamics, the scale parameter $R(t)$ depends on the initial
conditions as well as on  the equation of state, while in cosmology the time evolution 
of the scale parameter $R$
depends not only on the initial conditions (flat, open or closed universe)
and the  properties of matter (matter dominated and radiation dominated epoch)
but also on the possible existence of dark energy and cosmological constants, and the
possible presence of an exponentially accelerating, inflatory period.       
At the end of the accelerating period, when $\dot R =\,\hbox{const}$,	
and  $R \approx \dot R \, t$, the Hubble constant is simply the 
inverse of the lifetime, $H = 1/ t$. 

At  the very end of fireball explosions, the pressure decreases to vanishing values,
hence the acceleration caused by pressure gradients becomes
negligibly small.  In this case, the Hubble law 
connects the hydrodynamic  
flow four-velocity $u^\mu$ with the space-time position of the fluid element $x^\mu$,
in a simple way,
\begin{equation}
u^\mu = \frac{x^\mu}{\tau} = \frac{t}{\tau} \left(1, 
\frac{x}{t}, \frac{y}{t}, \frac{z}{t}\right).
\label{hubflow}
\end{equation}
The quantity $\tau$ in Eq. (\ref{hubflow}) is the proper time characterizing the 
freeze-out hypersurface, 
\begin{equation}
\tau = \sqrt{t^2 - r^2 - z^2}, \quad r = \sqrt{x^2 + y^2}.
\label{tau}
\end{equation}
Clearly, the parameterization (\ref{hubflow}) - (\ref{tau}) makes sense in the
space-time region defined by the condition $r^2+z^2<t^2$.
The three-velocity field of the form 
\begin{equation}
{\bf v} = \left( \frac{x}{t}, \frac{y}{t}, \frac{z}{t}\right),
\label{v}
\end{equation}
following directly from Eq. (\ref{hubflow}), appears very often in the analysis of 
the hydrodynamic equations applied to describe hadronic or nuclear collisions. In
this context it is called the asymptotic {\it scaling solution} \cite{Landau:1953gs,Belenkij:1956cd,
Cooper:1974ak,Cooper:1975qi,Bjorken:1983qr,Baym:1983sr}. In particular, for the 
boost-invariant systems the longitudinal velocity must be of the form $v_z = z/t$, 
which is a direct consequence of the Lorentz symmetry \cite{Hwa:1974gn,Chiu:1975hx,
Chiu:1975hw,Gorenstein:1977xv,Gorenstein:1978jg,Bjorken:1983qr}. 
It is often believed, that during the pre-equilibrium period of
high energy nuclear collisions, no significant transverse flow is generated
(for recent reviews of the hydrodynamic description of relativistic heavy-ion 
collisions see, e.g., Refs. \cite{Teaney:2001av,
Huovinen:2002fp,Kolb:2003dz,Hirano:2004er}).
In such a case, the transverse fluid velocity builds 
up during the hydrodynamic evolution of the system 
\cite{Baym:1983sr} and the scaling form $v_r = r/t$ may be reached only for 
sufficiently large times (with details depending on the equation of state and initial
conditions). Similar features characterize also a spherically symmetric 
expansion of the system being initially at rest. In this case, the radial flow 
is formed by the outward action of the pressure gradient, and the numerical
calculations show that Eq. (\ref{hubflow}) is the asymptotic solution 
\cite{Cooper:1975qi,Baym:1983sr}. 

Recently, there is a revived theoretical interest in finding exact solutions
of relativistic hydrodynamics. B\'{\i}r\'o found an interesting solution,
relevant for the case of a vanishing speed of sound, which interpolates
between an early Bjorken type of the flow profile and the Hubble profile in
the late period of the expansion~\cite{Biro:1999eh,Biro:2000nj}.
New exact solutions of relativistic hydrodynamics were found, using
more general equations of state, in the 1+1 dimensional case~\cite{Csorgo:2002ki} 
and in the 1+3 dimensional case; for axially symmetric~\cite{Csorgo:2002bi,Csorgo:2003rt},
as well as for ellipsoidally symmetric expansions~\cite{Csorgo:2003ry}.
Although these solutions contain arbitrary scaling functions
describing the rapidity profile, the flow profile in all of
these works coincides with the Hubble law.

In view of the success of the fits \cite{Broniowski:2001we,Csanad:2004mm,Retiere:2003kf,
Retiere:2004wa} which all indicate very strong transverse flows, one of the 
central issues is whether the times actually available in relativistic heavy-ion 
collisions are sufficient to allow for a dynamical development of such 
strong transverse flows, in particular, the scaling solutions corresponding to Eq. 
(\ref{hubflow}). The situation is especially intriguing for RHIC, where several 
measurements indicate an unexpected, rather short (about 10 - 15 fm) duration time of the 
collision process. For example, one finds $\tau \sim$ 10 fm using the RHIC data in the 
relation $R_L(m_T) = \tau \sqrt{T_k/m_T}$ \cite{Makhlin:1987gm}, which
connects the longitudinal pion correlation radius $R_L$, the kinetic freeze-out 
temperature $T_k$, the evolution time $\tau$, and the transverse mass of the pion pair $m_T$.

These measurements, when interpreted with care,
yield only the inverse of the (longitudinal) Hubble constant,
which can be identified with the lifetime only if the scaling solution
is assumed to be developed in the longitudinal direction.
This situation is analogous to the estimate of the lifetime of
the Universe: the inverse of the presently measured value of the Hubble
constant yields an order-of-magnitude estimate of the lifetime, which has to
be corrected for the effects of inflation and possible other acceleration
periods. Similarly, in nuclear hydrodynamics, there is  
an initial longitudinal acceleration period, hence the estimated 
lifetimes should be interpreted only as (lower) limits on the total lifetime of
the system~\cite{Csanad:2004mm}. 

Many hydrodynamic codes used to describe the heavy-ion data
show that the scaling solutions do not appear before the freeze-out of the system. 
However, such approaches assume commonly that the initial transverse flow is
zero. An exception from this rule is the work by Kolb and Rapp \cite{Kolb:2002ve},
where the pre-equilibrium transverse flow is considered and its presence improves
the agreement of the model calculations with the data.
Another important exception is a class of the non-relativistic, selfsimilar
solutions of the fireball hydrodynamics, which is by now solved completely
in the ellipsoidally symmetric case for abritrary initial sizes and expansion
velocities of the principal axes of the expanding ellipsoids~\cite{Akkelin:2000ex}, 
arbitrary initial temperature profiles~\cite{Csorgo:2001ru},
as well as for arbitrary (temperature dependent) speed of sound~\cite{Csorgo:2001xm}.

In this paper we follow such ideas and assume that the elementary parton-parton collisions, 
leading to the thermalization of the system, lead also to collective behavior and development 
of the transverse flow already at the initial stage of the equilibrium hydrodynamic evolution 
at the time $t = t_0 \sim 1$ fm. For simplicity, we consider the boost-invariant and 
cylindrically symmetric systems with the initial transverse flow defined by the formula
\begin{equation}
v^{\,0}_r =   v_r(t=t_0,r) = \frac{ H r}{\sqrt{1 + H^2 r^2}}.
\label{initvr}
\end{equation}
The parameter $H$ in Eq. (\ref{initvr}) may be interpreted as the Hubble constant which 
determines the magnitude of the initial transverse flow; in the range $r < 1/H$ the flow
is well approximated by the linear function $v^{\,0}_r \sim H r$, whereas for
$r > 1/H$ the flow approaches the speed of light, a boundary condition 
frequently assumed in the 
hydrodynamic equations for large values of $r$ \cite{Baym:1983sr}. Also, if the
above equation is specified on a hypersurface corresponding to a constant
proper-time, it coincides with the Hubble law of Eq.~(\ref{hubflow}).

By varying the value of $H$ we control the amount of the initial transverse flow which may 
possibly develop into the scaling form (\ref{v}). We note that in the scaling solution 
(\ref{v}) the role of the 
Hubble constant is played by the inverse of the time coordinate $t$. This means that setting
$H = 1/t_0 =1 \hbox{fm}^{-1}$ we assume that the initial flow is already very 
close to the scaling 
form at $t_0$. In this particular case the question arises if the flow remains close to the 
scaling form in the subsequent time evolution of the system.

A few comments are in order now. Following the conventional hydrodynamic approach 
\cite{Baym:1983sr}, we specify the initial conditions at a constant time, $t_0 = 1$ fm, 
and search for the solutions which are regular functions of $t$ and $r$ in the region: 
$t > t_0$, $r > 0$. In this case, at a given value of time $t$, the scaling solution 
cannot hold for arbitrary large values of $r$, since this would yield the flow velocities 
larger than the speed of light,
and also the initial condition deviates from Eq. (1) for large values of the 
transverse radius. Hence, in our analysis 
we search for the solutions of the hydrodynamic equations which yield the flow profiles 
possibly close to the scaling solution in the region $r < t$.
In a separate paper we intend to explore a different type of the initial conditions, 
which are specified at a fixed value of the proper time and lead, in certain special cases, 
to the exact solutions given in Refs.~\cite{Csorgo:2003rt,Csorgo:2003ry}.
We also note that in view of the recent BRAHMS data \cite{Bearden:2003fw,Bearden:2004yx} 
describing rapidity dependence of hadron production, the
boost-invariant approach assumed 
in the present calculation may be appropriate only if limited to the
rapidity range: $-1 < y < 1$.

\section{Characteristic form of the hydrodynamic equations}

In this Section we introduce the basic notation and rewrite the hydrodynamic equations 
in the form convenient for the numerical calculations. We follow closely the method 
introduced by Baym, Friman, Blaizot, Soyeur, and Czy\.z \cite{Baym:1983sr}. We restrict 
our considerations to the systems with zero net baryon density, which is a good
approximation for description of the central rapidity region at the RHIC energies 
(thermal analysis of the ratios of hadron multiplicities indicates that the baryon
chemical potential at RHIC energies is about 30 MeV, i.e., it is much smaller
than the corresponding temperature of about 170 MeV \cite{Florkowski:2001fp,
Torrieri:2004zz,Csanad:2004mm}). In this case the hydrodynamic equations have the form
\begin{equation}
u^{\mu }\partial _{\mu }\left( T\,u^{\nu }\right) =\partial ^{\nu }T,
\label{acc}
\end{equation}
\begin{equation}
\partial _{\mu }\left( \sigma u^{\mu }\right) =0,
\label{ent}
\end{equation}
where $T$ is the temperature, $\sigma $ is the entropy density, and $u^{\mu
}=\gamma \left( 1,\mathbf{v}\right) $ is the hydrodynamic four-velocity
(with $\gamma =1/\sqrt{1-v^{2}}$).  Eq. (\ref{acc}) is the acceleration equation
which is the analog of the Euler equation of \ the classical hydrodynamics, whereas
Eq. (\ref{ent}) represents entropy conservation (the adiabaticity of the flow).
Since $T$ is the only independent
thermodynamic variable, all other thermodynamic quantities can be obtained
from the equation of state $P\left( T\right)$, connecting pressure $P$
with the temperature $T$. With the help of the thermodynamic relations
\begin{equation}
d\varepsilon =Td\sigma ,\qquad dP=\sigma dT,\qquad w=\varepsilon +P=T\sigma,
\label{thermo}
\end{equation}
other thermodynamic quantities, such as the energy density $\varepsilon $ or
the enthalpy density $w$, can be obtained. In addition, the equation of
state allows us to calculate the sound velocity
\begin{equation}
c_{s}^{2}=\frac{\partial P}{\partial \varepsilon }=\frac{\sigma }{T}\frac{%
\partial T}{\partial \sigma }.
\label{cs}
\end{equation}
Equation (\ref{ent}) and the spatial components of Eq. (\ref{acc}) may be 
rewritten
for the boost invariant systems with cylindrical symmetry as
\begin{eqnarray}
v_{r}\frac{\partial \ln T}{\partial t}+\frac{\partial \ln T}{\partial r}+%
\frac{\partial \alpha }{\partial t}+v_{r}\frac{\partial \alpha }{\partial r}
 & = &0, \label{accf}  \\
\frac{\partial \ln \sigma }{\partial t}+v_{r}\frac{\partial \ln \sigma }{%
\partial r}+v_{r}\frac{\partial \alpha }{\partial t}+\frac{\partial \alpha }{%
\partial r}+\frac{1}{t}+\frac{v_{r}}{r} & = &0,
\label{entf}
\end{eqnarray}
where $\alpha$ is the transverse rapidity of the fluid defined by
the condition $v_{r}=\tanh\alpha$. The longitudinal
component has the well known boost-invariant form $v_{z}=z/t$ \cite{Bjorken:1983qr}.

By introducing  the potential $\Phi \left( T\right) $ defined by the
differentials
\begin{equation}
d\Phi =\frac{d\ln T}{c_{s}}=c_{s}d\ln \sigma ,
\label{phi}
\end{equation}
and by the use of  the two functions $a_{\pm }$ defined by the formula
\begin{equation}
a_{\pm }=\exp \left( \Phi \pm \alpha \right),
\label{apm}
\end{equation}
Eqs. (\ref{accf}) and (\ref{entf}) may be cast into the characteristic form
\cite{Baym:1983sr}
\begin{eqnarray}
\frac{\partial }{\partial t}a_{\pm }\left( t,r\right) +\frac{v_{r}\pm c_{s}}
{1\pm v_{r}\,c_{s}}\frac{\partial }{\partial r}\,a_{\pm }\left( t,r\right) 
+ \frac{c_{s}}{1\pm v_{r}\,c_{s}}\left( \frac{v_{r}}{r}+\frac{1}{t}\right)
a_{\pm }\left( t,r\right) = 0.
\label{eqapm}
\end{eqnarray}
If the functions $a_{\pm}$ are known, the potential $\Phi$ may
be calculated from the formula
\begin{equation}
\Phi = \frac{1}{2}\ln \,\left( a_{+} a_{-}\right),
\label{phioft}
\end{equation}
and the velocity is obtained from the equation
\begin{equation}
v_{r}  =\frac{a_{+} -a_{-}}{a_{+}+ a{-}}.
\label{vofa}
\end{equation}
The knowledge of the function $c_s(T)$ allows us, by the integration of 
Eq. (\ref{phi}), to determine $\Phi$ as a function of the temperature; 
this function will be called later $\Phi_T$. However, to get a closed system of 
equations for $a_+$ and $a_-$, we have to invert this relation and obtain $T$ as 
a function of $\Phi$; this function will be called later $T_\Phi$. In this way, 
the sound velocity may be expressed in terms of the functions $a_+$ and $a_-$,
\begin{equation}
c_s=c_s\left[T_\Phi\left(\frac{1}{2}\ln \,\left( a_{+} a_{-}\right) 
\right)\right],
\label{csapam}
\end{equation}
and Eqs. (\ref{eqapm}) may be solved numerically.

\section{Temperature dependent sound velocity}

\begin{figure}[t]
\begin{center}
\subfigure{\includegraphics[angle=0,width=0.45\textwidth]{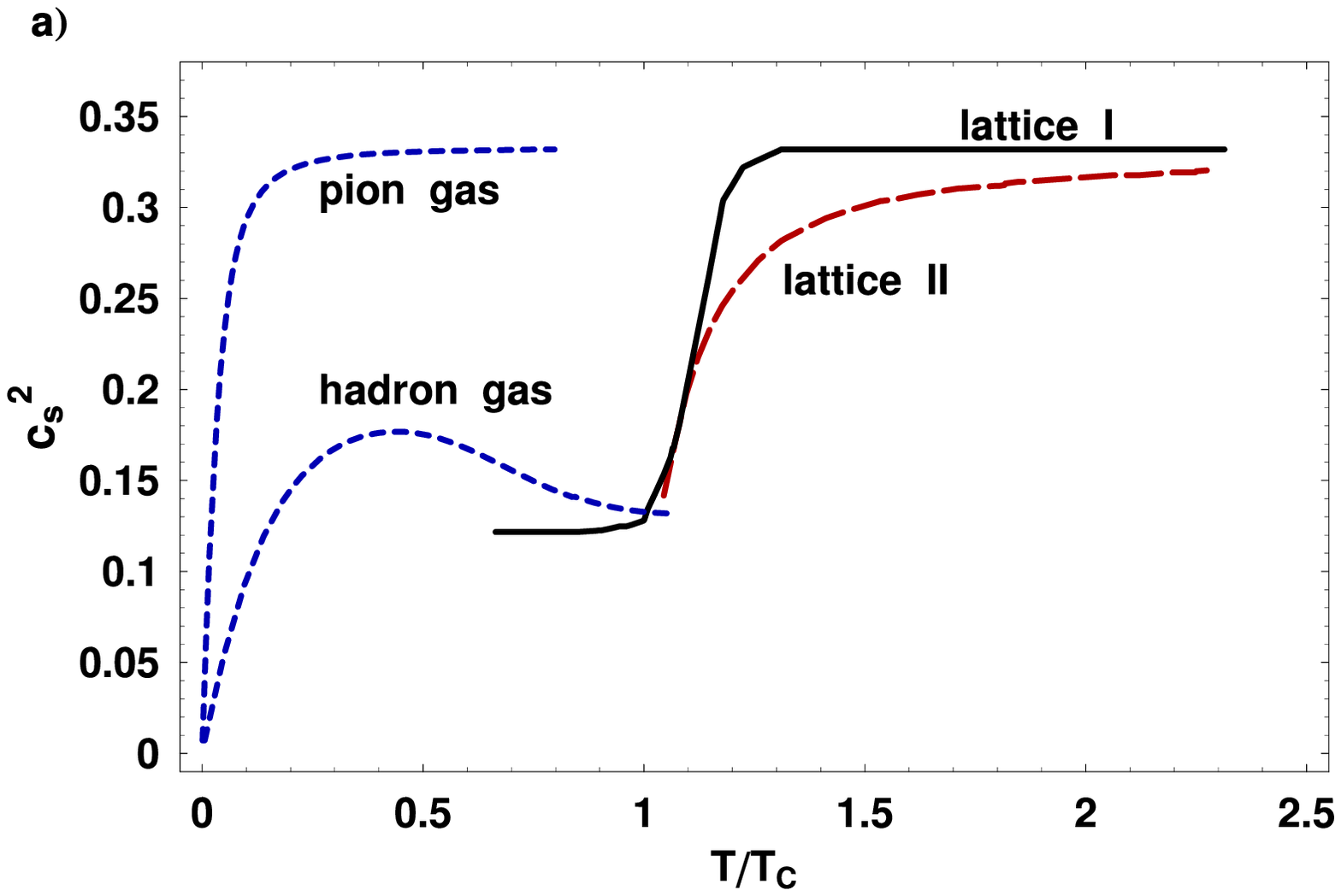}} 
\subfigure{\includegraphics[angle=0,width=0.45\textwidth]{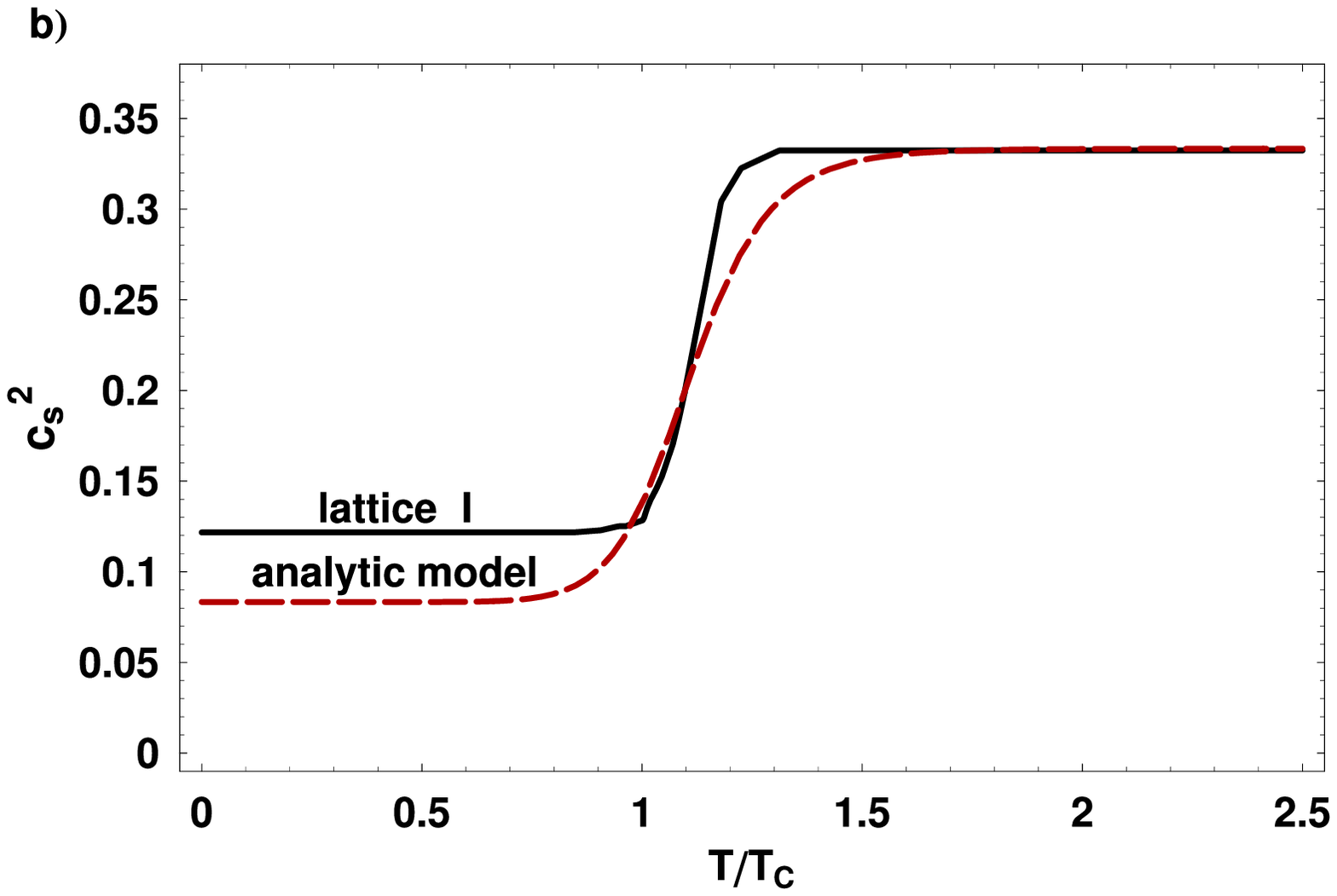}} \\
\end{center}
\caption{ {\bf a)} Temperature dependence of the sound velocity as obtained 
from different theoretical models described in the text. {\bf b)}  
Sound velocity used in this calculation: the solid line describes the lattice result, 
i.e., the function "lattice I" from part a) extrapolated to low temperatures, whereas
the dashed line describes the analytic model defined by Eq. (\ref{toycs}).}
\label{fig:cs}
\end{figure}
In our numerical calculations we take into account the temperature dependence of 
the sound velocity. In this way, we generalize the approach of Ref. \cite{Baym:1983sr}, 
where Eqs. (\ref{eqapm}) - (\ref{vofa}) were solved numerically only in the case 
$c^2_s = 1/3$. The results of different model calculations of the sound velocity, 
which may serve as the input for the hydrodynamic calculations are presented in 
Fig. \ref{fig:cs} a). The solid line (denoted as "lattice I") is the result of Mahonty and 
Alam \cite{Mohanty:2003va}, who compiled the lattice results obtained by
Karsch \cite{Karsch:2001vs} in order to get $c_s(T)$ from the temperature 
dependence of the energy density. The long-dashed line (denoted as "lattice II")
shows the result of the lattice QCD calculations by Szab\'o and T\'oth \cite{Szabo:2003kg}.

A sudden but smooth change of the sound velocity in the small temperature range around 
$T=T_c$, as observed in the lattice calculations, see Fig. \ref{fig:cs} a), indicates 
a rapid cross-over from a hadron gas to a quark-gluon plasma phase. 
Above the critical temperature, ($T \gg T_c$) the sound 
velocity approaches the limit valid for massless particles, $c^2_s = 1/3$, 
whereas below the 
phase transition ($T \ll T_c$) the sound velocity is much smaller, 
being close to the value 
obtained  for the case of the noninteracting gas of hadron resonances.

The hadron-gas result shown in Fig. \ref{fig:cs} a) was obtained by us in
the calculation which uses the recent fits to the meson and baryon mass 
spectra \cite{Broniowski:2000bj,Broniowski:2000hd,Broniowski:2004yh} denoted below by 
$\rho_M(m)$ and $\rho_B(m)$. One possible parameterization of the spectra 
\cite{Broniowski:2000hd}, which reveals directly 
their exponential growth, as suggested long ago by Hagedorn
\cite{Hagedorn:1965st}, is as follows

\begin{eqnarray}
\rho_M(m) &=& a_M \exp(m/T_M), \quad a_M = 4.41 \, \hbox{GeV}^{-1}, \quad
T_M = 311 \,\hbox{MeV},  \nonumber \\
\rho_B(m) &=& a_B \exp(m/T_B), \quad a_B = 0.11 \, \hbox{GeV}^{-1}, \quad 
T_B = 186 \,\hbox{MeV}.
\label{tmtb}
\end{eqnarray}
With the help of the mass distributions (\ref{tmtb}), we 
calculate the entropy density of the hadron gas as a sum over contributions
from all known hadronic states from the formula

\begin{eqnarray}
\sigma(T) 
= \frac{1}{2 \pi^2} \int_{m_{\rm pion}}^{M_{\rm mesons}^{\rm max}} \rho_M(m)  
m^3 K_3 \left( \frac{m}{T} \right) dm  
+ \frac{2}{2 \pi^2} \int_{m_{\rm nucleon}}^{M_{\rm baryons}^{\rm max}} \rho_B(m) 
m^3 K_3 \left( \frac{m}{T} \right) dm. 
\label{clasent}
\end{eqnarray}
The sound velocity of the hadron gas follows then directly from the use of
Eq. (\ref{clasent}) in (\ref{cs}). Eq. (\ref{clasent}) is valid in the case of zero baryon 
chemical potential and the factor 2 in the second term accounts for antibaryons. The effects 
of the quantum statistics (Bose-Einstein or Fermi-Dirac) are neglected in Eqs.~(\ref{clasent}), 
because they are known to be small; of the order of 20\% or less for reasonable range of the 
temperatures~\cite{Belenkij:1956cd}. To match our hadron-gas calculation with the lattice 
data we assumed that the critical temperature is 170 MeV. The upper limits of the 
integrations in (\ref{clasent}) are $M^{\rm max}_{\rm mesons}$ = 2.3 GeV for 
mesons and $M^{\rm max}_{\rm baryons}$ = 1.8 GeV for baryons. These limits are
determined by the range where the fit (\ref{tmtb}) works well
\cite{Broniowski:2004yh}; for higher masses, 
due to the lack of data, the spectra saturate. 

For comparison, in Fig. \ref{fig:cs} a)
we show the speed of sound obtained in the similar calculation where the massive pions 
are included only. The speed of sound in the pion gas reaches very fast 
the limiting value of $1/\sqrt{3}$, which may be confronted with the non-monotonic
behavior of $c_s$ in the gas of resonances. Similar  behavior of the speed of sound in 
the gas of resonances was found also in the case with non-zero baryon
density \cite{Prorok:2002ht}. It is interesting to note that the lattice data agree with
the hadron-gas calculation close to the phase transition region 
if we assume $T_c \sim$ 170 MeV. Moreover, the speed of sound in the hadron resonance
gas below the critical temperature is found to be significantly smaller than
$1/\sqrt{3}$, which is the limit of massless ideal relativistic gas used
in the bag model type of the equations of state. As the hydrodynamic
equations describe the properties of matter through the equation of
state, or more precisely through the speed of sound~\cite{Belenkij:1956cd},
such decrease of the speed of sound in the hadron gas stage, compared to the
massless pion gas limit, changes drastically the corresponding 
time evolution of the hydrodynamic solutions.

\begin{figure}[t]
\begin{center}
\subfigure{\includegraphics[angle=0,width=0.45\textwidth]{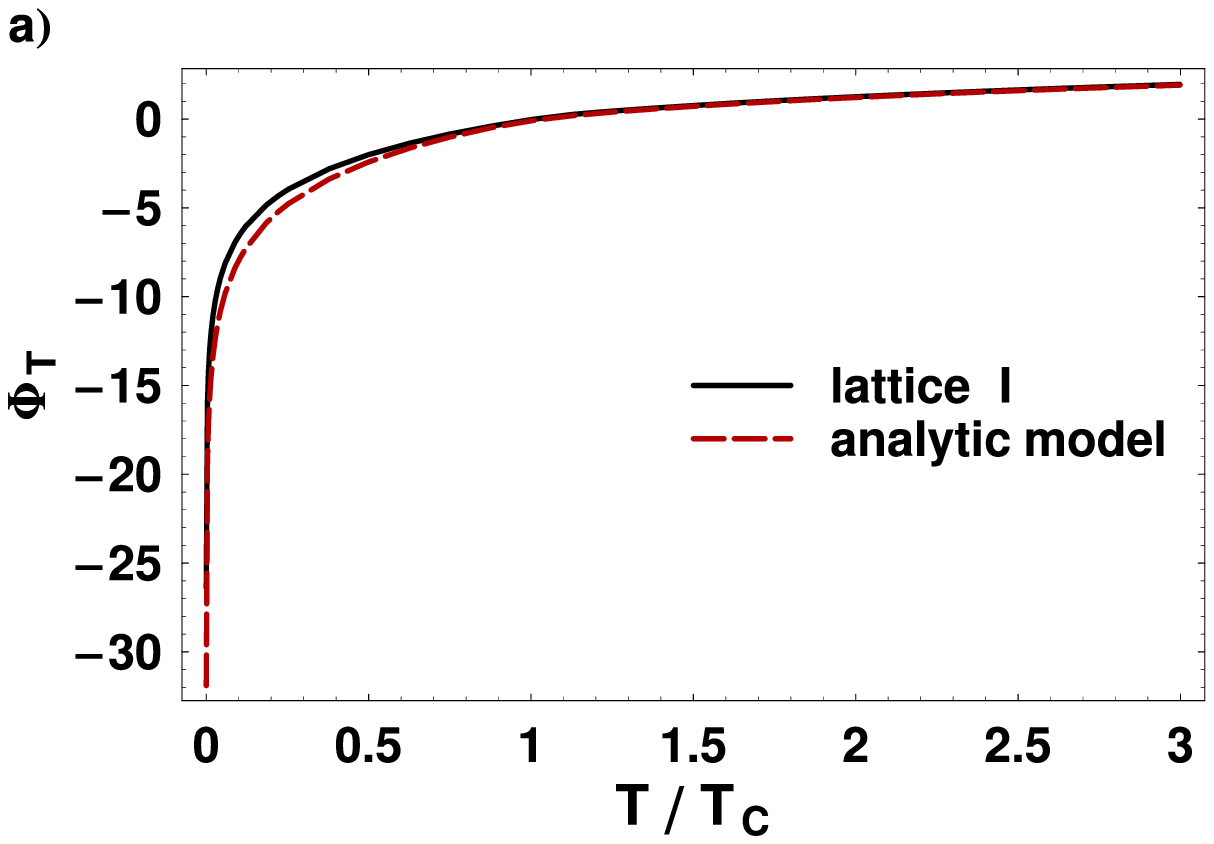}} 
\subfigure{\includegraphics[angle=0,width=0.45\textwidth]{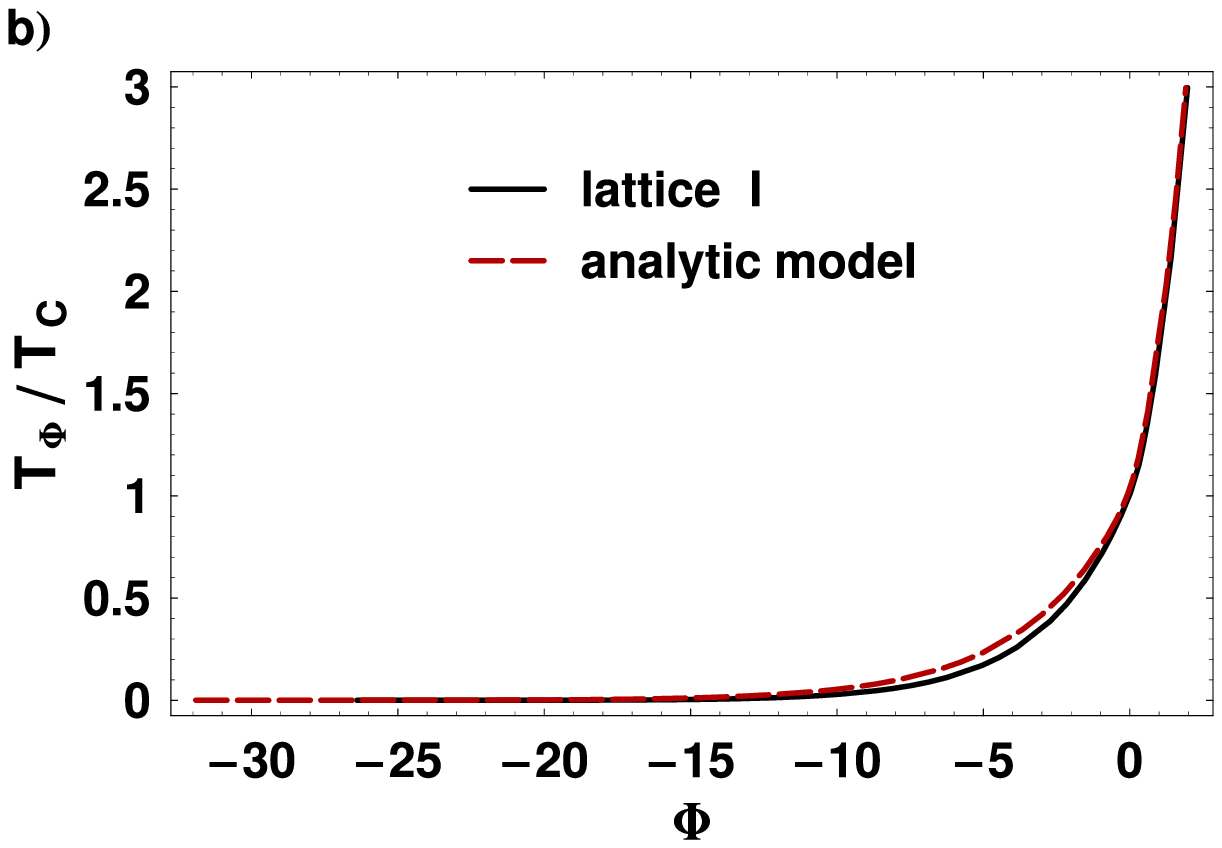}} \\
\end{center}
\caption{ {\bf a)} The potential $\Phi$ obtained
by the integration of Eq. (\ref{phi}). {\bf b)} Temperature
as a function of the potential $\Phi$. In both cases, a) and b), the
solid lines show the results obtained with the lattice equation
of state, whereas the dashed lines describe the analytic model
defined by Eq. (\ref{toycs}).}
\label{fig:phit}
\end{figure}

Although the calculations of the speed of sound shown in Fig. \ref{fig:cs} a) are based 
on the observed hadron mass spectra below $T_c$ and on the lattice QCD above $T_c$, they are 
still somewhat ambiguous and not completely satisfactory. For example, lattice QCD calculations 
below the critical temperature still yield too heavy pions, and we know that the value of the 
speed of sound is rather sensitive to the pion mass. On the other hand, the calculation based on 
the Hagedorn mass spectrum in the hadronic phase is more realistic than the lattice results in 
the low temperature domain, however, it neglects the role of interactions between the hadrons.

In this situation, we decided to use as the main input the equation of state (sound velocity)
delivered by Ref. \cite{Mohanty:2003va}, which is shown as "lattice I" in Fig. \ref{fig:cs} a).
This calculation is based on first principles and extends down to about 0.6 $T_c$, i.e., to the 
low-temperature region of about 100 MeV relevant for the discussions of the kinetic
freeze-out. In the actual calculation we extrapolate this result to even lower temperatures as
shown in Fig. \ref{fig:cs} b). We also use an analytic form of the function $c_s(T)$ which 
exhibits the main features observed in Fig. \ref{fig:cs} a) and, at the same time, leads to 
the analytic expressions for the functions $\Phi_T(T)$ and $T_\Phi(\Phi)$.
This function is defined by the formula
\begin{equation}
c_s(T) = \frac{1}{\sqrt{3}} 
\left[1 - \frac{1}{2}
\left(
\frac{1}{1 + (T/{\tilde T})^{2 n}}
\right) \right].
\label{toycs}
\end{equation}
Using Eq. (\ref{toycs}) one finds
$c_s(T) = 1/\sqrt{3}$ for $T \gg {\tilde T}$ and $c_s(T) = 1/(2 \sqrt{3})$ 
for $T \ll {\tilde T}$. A straightforward integration of Eq. (\ref{phi}) gives 
in this case
\begin{equation}
\Phi_T(T) = \frac{\sqrt{3}}{ 2 n} \,
\hbox{ln} \,
\frac{(T/{\tilde T})^{4n}}{1+2 (T/{\tilde T})^{2n} }
\label{toyphi}
\end{equation}
and
\begin{equation}
T_\Phi(\Phi) = {\tilde T}
\left[
e^\frac{2 n \Phi }{\sqrt{3} } +
\sqrt{
e^\frac{2 n \Phi}{\sqrt{3} } 
+ e^\frac{4 n \Phi}{\sqrt{3} } }
\, \right]^\frac{1}{2 n}.
\label{toytemp}
\end{equation}
With the parameters ${\tilde T} = 1.08\,T_c$ and $n=6$ the function 
(\ref{toycs}) well approximates the lattice results I and II in the region slightly above
$T_c$,  and interpolates between the lattice results I and II at higher temperatures.
At low temperatures, $T < 0.6\,T_c$, the function (\ref{toycs}) behaves like
a constant, see Fig. \ref{fig:cs} b). The functions $\Phi_T(T)$ and $T_\Phi(\Phi)$,
defined by Eqs. (\ref{toyphi}) and (\ref{toytemp}) with ${\tilde T} = 1.08\,T_c$ 
and $n=6$ are represented in Fig. \ref{fig:phit} by the dashed lines. The solid
lines in Fig. \ref{fig:phit} show the same functions obtained for the lattice case.
({\it In the following, we shall refer to the lattice calculations having in mind
the case "lattice I",  including a linear extrapolation at very low temperatures}).

\section{Initial conditions }

From the symmetry reasons, the velocity field should vanish at $r=0$. This 
condition is achieved if the functions $a_+$ and $a_-$ are initially determined 
by a single function $a(r)$ according to the prescription  \cite{Baym:1983sr}
\begin{eqnarray}
a_+(t=t_0,r)=a(r), \quad a_-(t=t_0,r)=a(-r).
\label{inita}
\end{eqnarray}
In the following we shall assume that the hydrodynamic evolution starts at a typical
time $t=t_{0}=1$ fm.  We shall also assume that the initial temperature profile
is connected with the nucleon-nucleus thickness function $T_A(r)$,
\begin{equation}
T(t=t_0,r) = \hbox{const} \, T^{1/3}_A(r).
\label{T01}
\end{equation}
where
\begin{equation}
T_A(r) = 2 \int d^2 s \int dz \, \rho\left(\sqrt{({\bf s-r})^2+z^2}\right).
\label{TA}
\end{equation}
Here the function $\rho(r)$ is the nuclear density profile given by the
Woods-Saxon function with a conventional choice of the parameters
($\rho_0 = 0.17 \,\hbox{fm} ^{-3}, \, r_0 = (1.12 A^{1/3} -0.86 A^{-1/3}) \,\hbox{fm} ,
\, a = 0.54 \,\hbox{fm}, \, A = 197$).

The idea
to use Eq. (\ref{T01}) follows from the assumption that the initially produced
entropy density  $\sigma(r)$ is proportional to the number of the nucleons 
participating in the collision at a distance $r$ from the collision center
\cite{Kolb:2003dz}, 
$\sigma(r) \sim T_A(r)$. Since for massless particles the entropy density is 
proportional to the third power of the temperature, we arrive at Eq. (\ref{T01}).

We note that other choices for the initial temperature profile are also conceivable.
If we assume that the initially produced energy density is proportional
to the nuclear thickness function, instead of Eq. (\ref{T01}) we obtain 
\begin{equation}
T(t=t_0,r) = \hbox{const} \, T^{1/4}_A(r).
\label{T01b}
\end{equation}
If we assume that the energy deposition into the thermalization
is driven by the collisions of wounded nucleons, and every collision
contributes with certain probability distribution to a local increase of the temperature,
then after many collisions the central limit theorem may describe the
form of the local temperature distribution and this is a Gaussian form. 
However, if there are big fluctuations in the deposited energy,
the generalized central limit theorems apply and in this case
the local temperature distribution may have the generalized,
L\'evy stable form. As a special case, the Lorentzian temperature
profile can also be obtained.

The two initial conditions (\ref{initvr}) and (\ref{T01}) may be included in the
initial form of the function $a(r)$ if we define
\begin{equation}
a(t=t_0,r) =
a_T(r) \frac{\sqrt{1 + v^{\,0}_r}}{\sqrt{1 - v^{\,0}_r} },
\label{initaT}
\end{equation}
where
\begin{equation}
a_T(r) = \hbox{exp}
\left[\Phi_T
\left(
\hbox{const} \,\, T_A^\frac{1}{3}(|r|)
\right)
\right].
\label{aT}
\end{equation}

\section{Results}

\begin{figure}[t]
\begin{center}
\subfigure{\includegraphics[angle=0,width=0.4\textwidth]{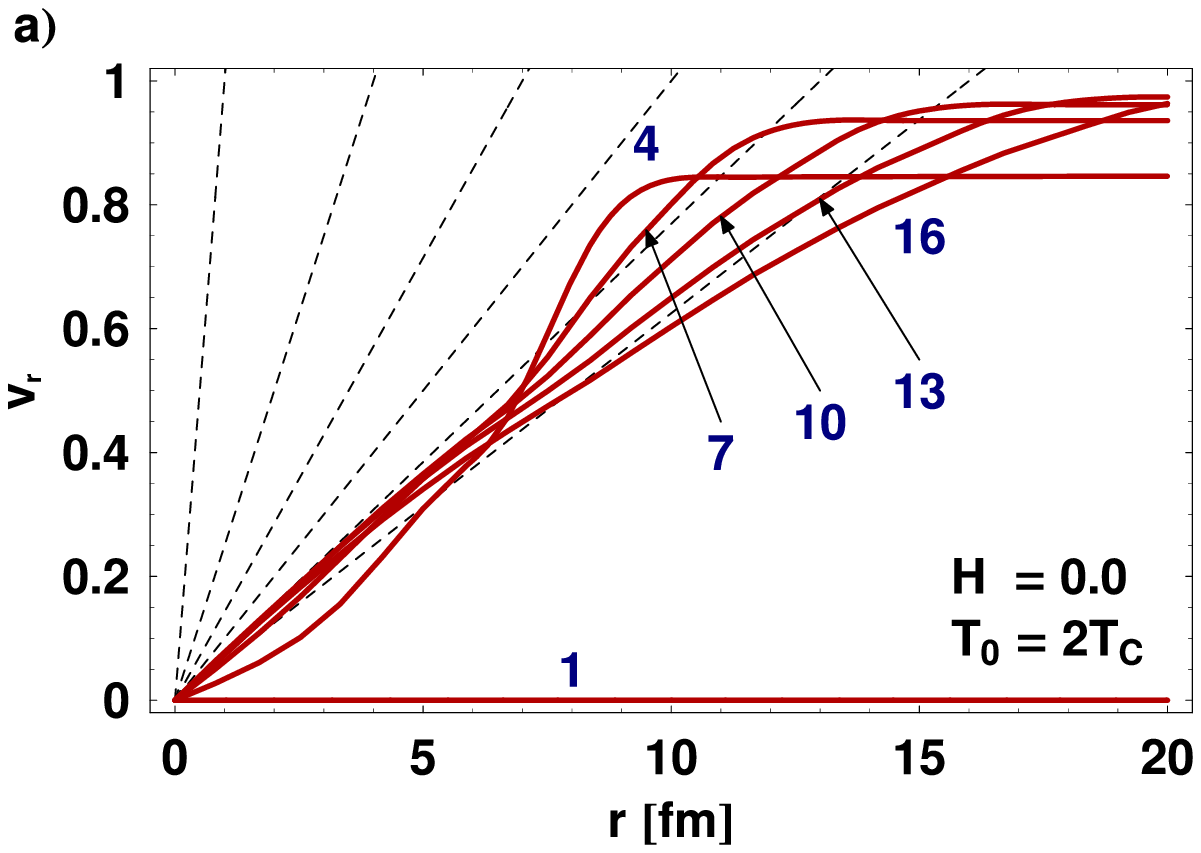}} 
\subfigure{\includegraphics[angle=0,width=0.4\textwidth]{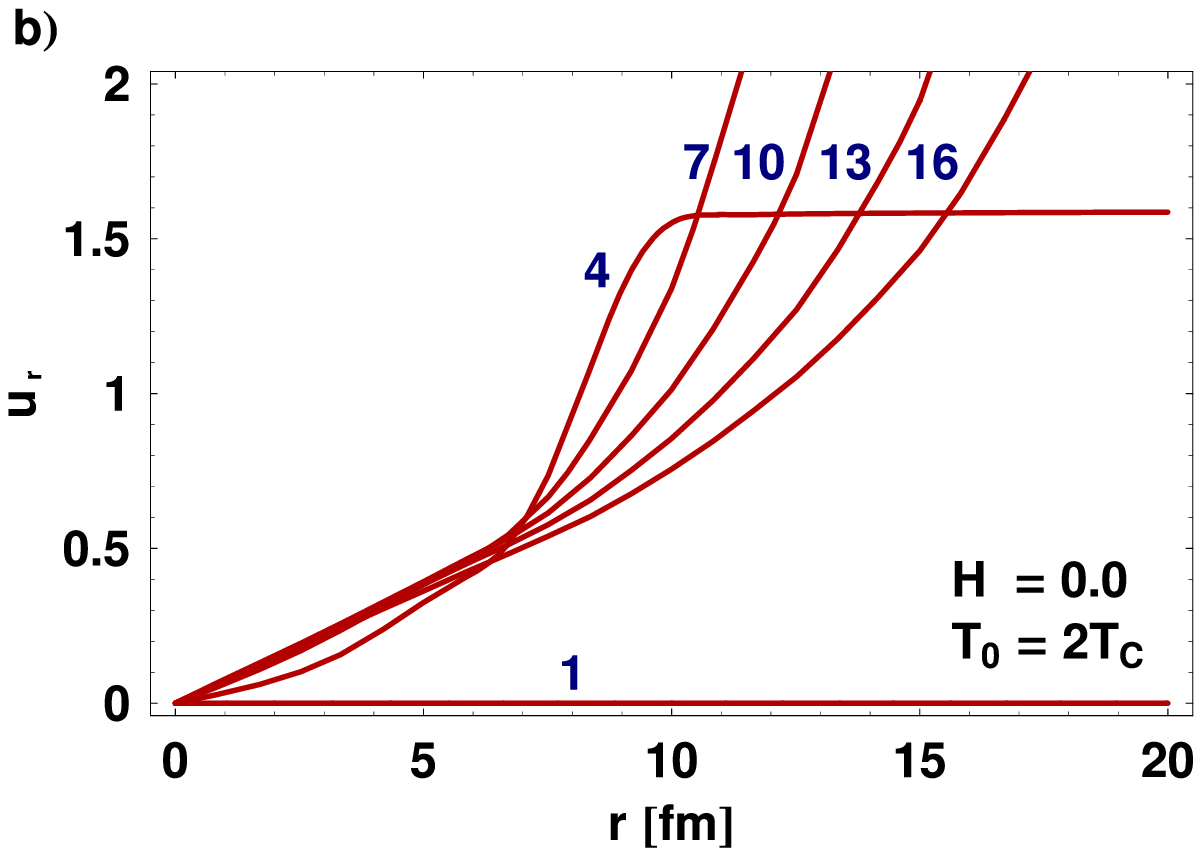}} \\
\subfigure{\includegraphics[angle=0,width=0.4\textwidth]{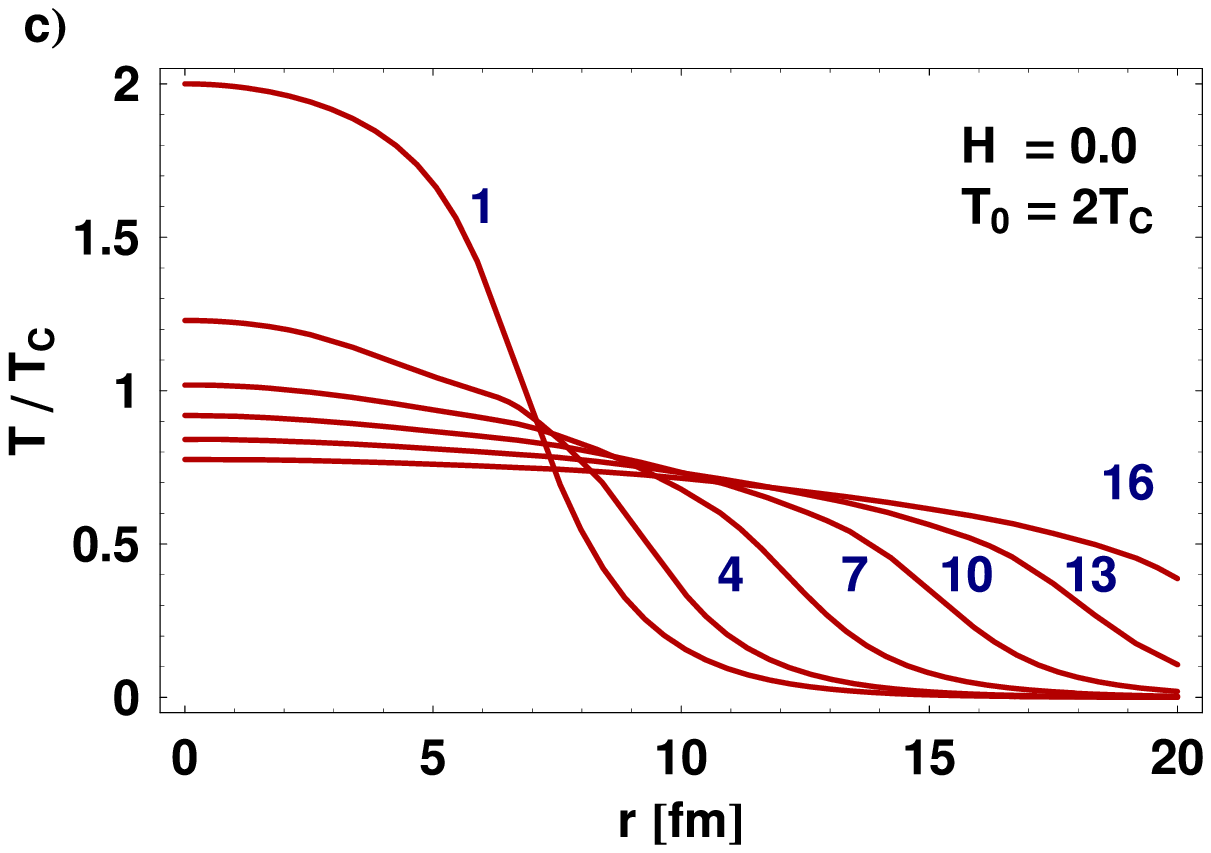}} 
\subfigure{\includegraphics[angle=0,width=0.4\textwidth]{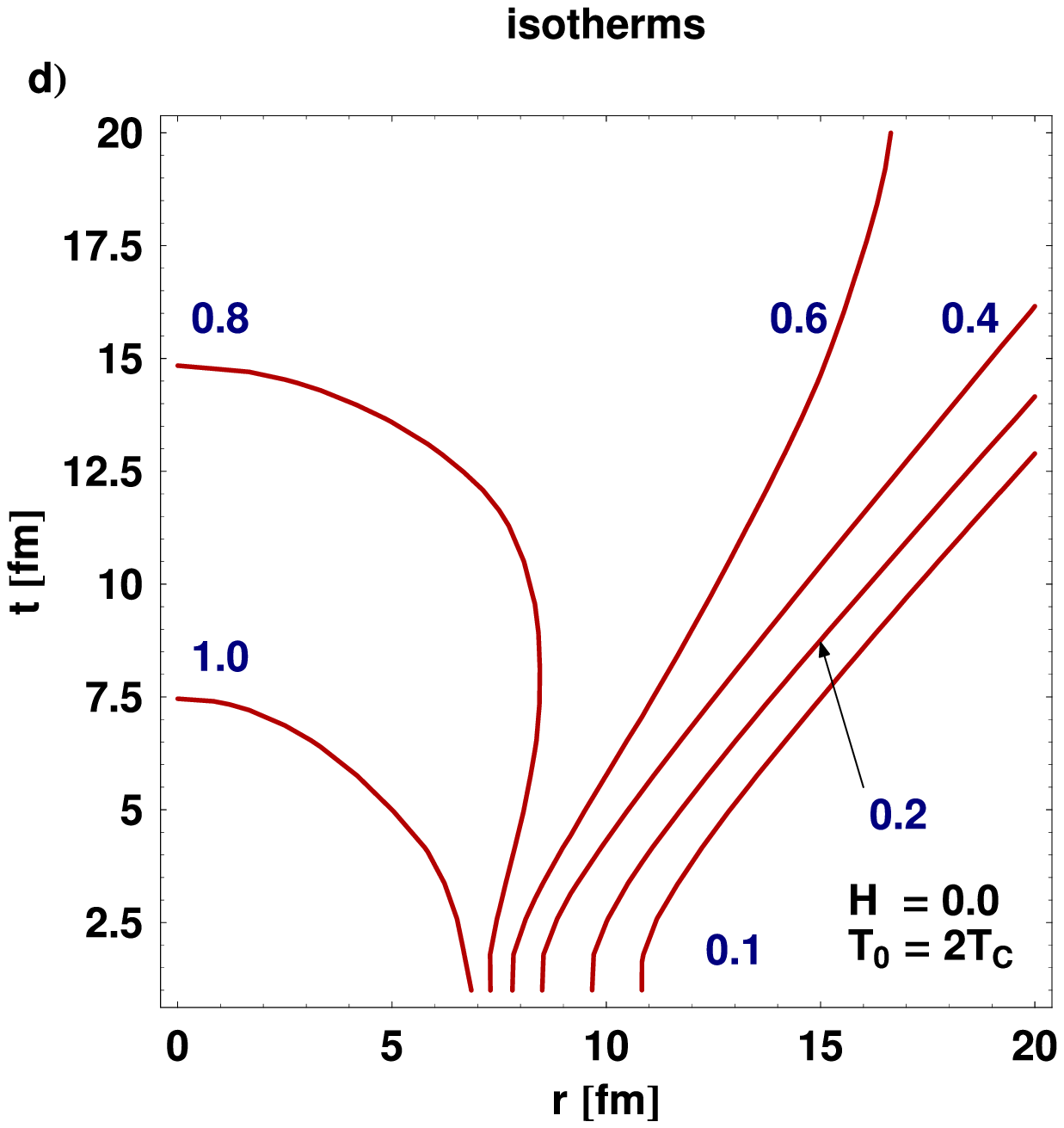}} \\
\end{center}
\caption{Hydrodynamic expansion of matter being initially at rest, i.e., in the
case $H=0$. The initial central temperature $T_0 = 2 T_c$.
The part {\bf a)} shows the transverse velocity as a function of the 
distance from the center for 6 different values of time, 
$t_i$\,=\,1,\,4,\,7,\,10,\,13 and 16 fm. The dashed thin lines describe the
ideal Hubble-like profiles of the form $r/t_i \, (r < t_i)$.
The part {\bf b)} shows the transverse four-velocity as a function of $r$ for the same
values of time. The part {\bf c)} shows the temperature profiles in $r$ 
whereas the part {\bf d)} shows the isotherms in the 
$t-r$ space.  In this case, the labels at the curves denote the 
values of the temperature in units of the critical temperature.}
\label{fig:HIN2TH00}
\end{figure}

It is important to observe, that the Hubble law $v = H r$ is satisfied
in the $r \ll t$ region after $t = 7 \,$ fm in all the cases
that we explored numerically in the present calculation, 
regardless of the initial conditions. However, the value of the Hubble constant
typically deviates form the inverse of time, which signals that the
asymptotic form of the Hubble flow, eq.~(\ref{hubflow}) is reached only
after a longer time period. The onset of the asymptotic scaling is
determined by the initial conditions as we shall detail below.

\begin{figure}[t]
\begin{center}
\subfigure{\includegraphics[angle=0,width=0.4\textwidth]{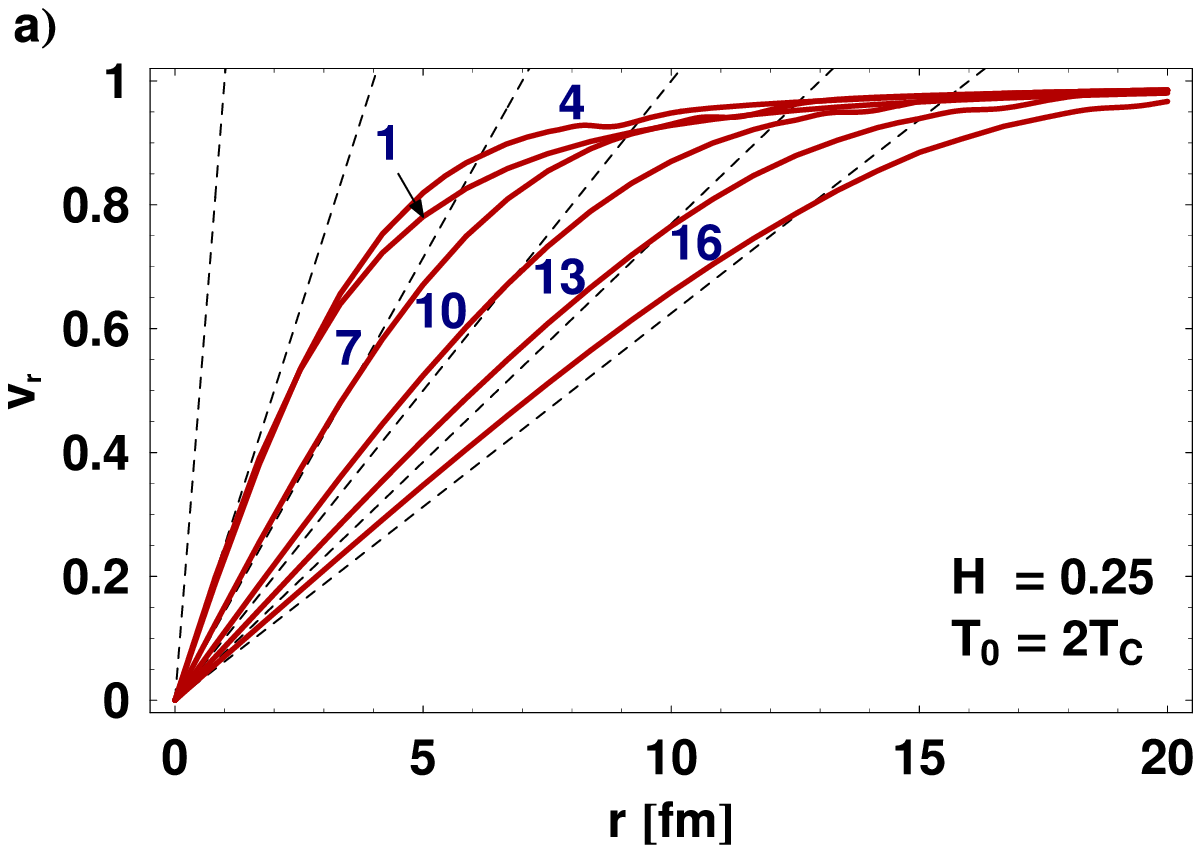}} 
\subfigure{\includegraphics[angle=0,width=0.4\textwidth]{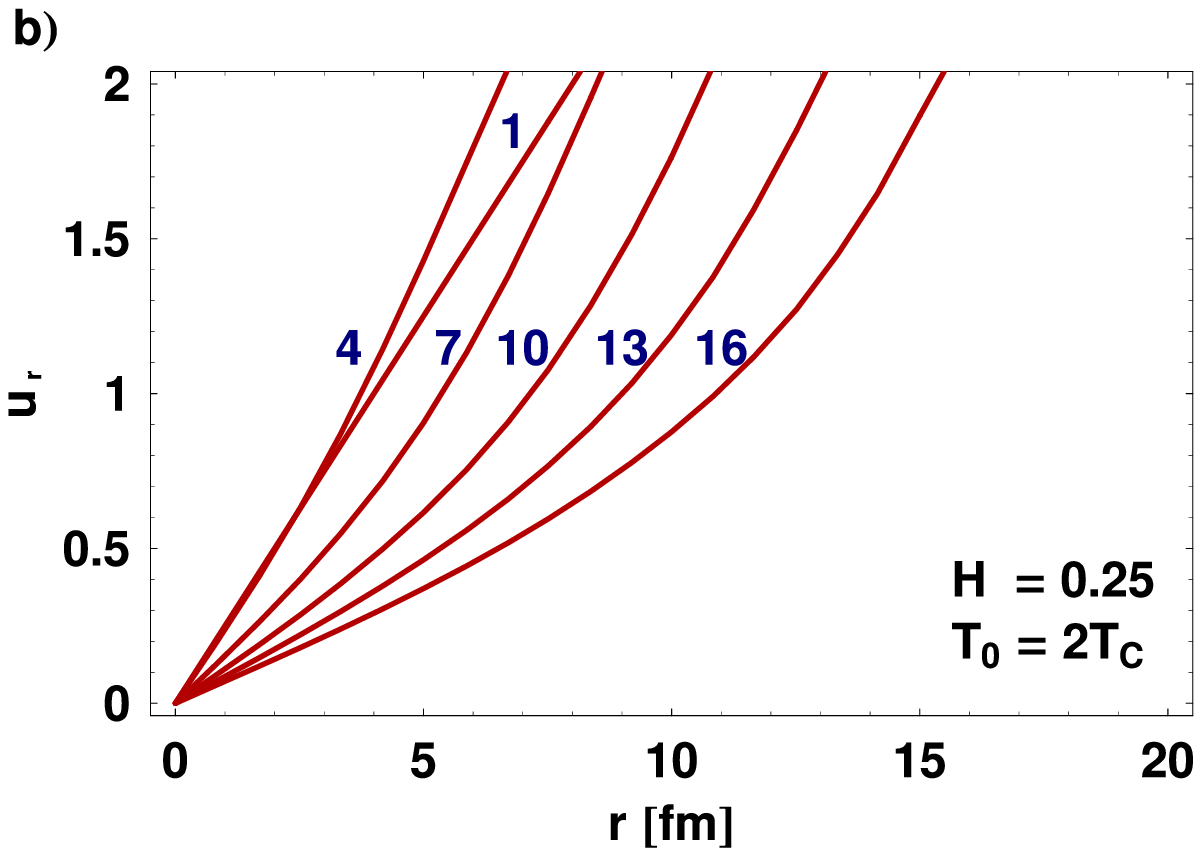}} \\
\subfigure{\includegraphics[angle=0,width=0.4\textwidth]{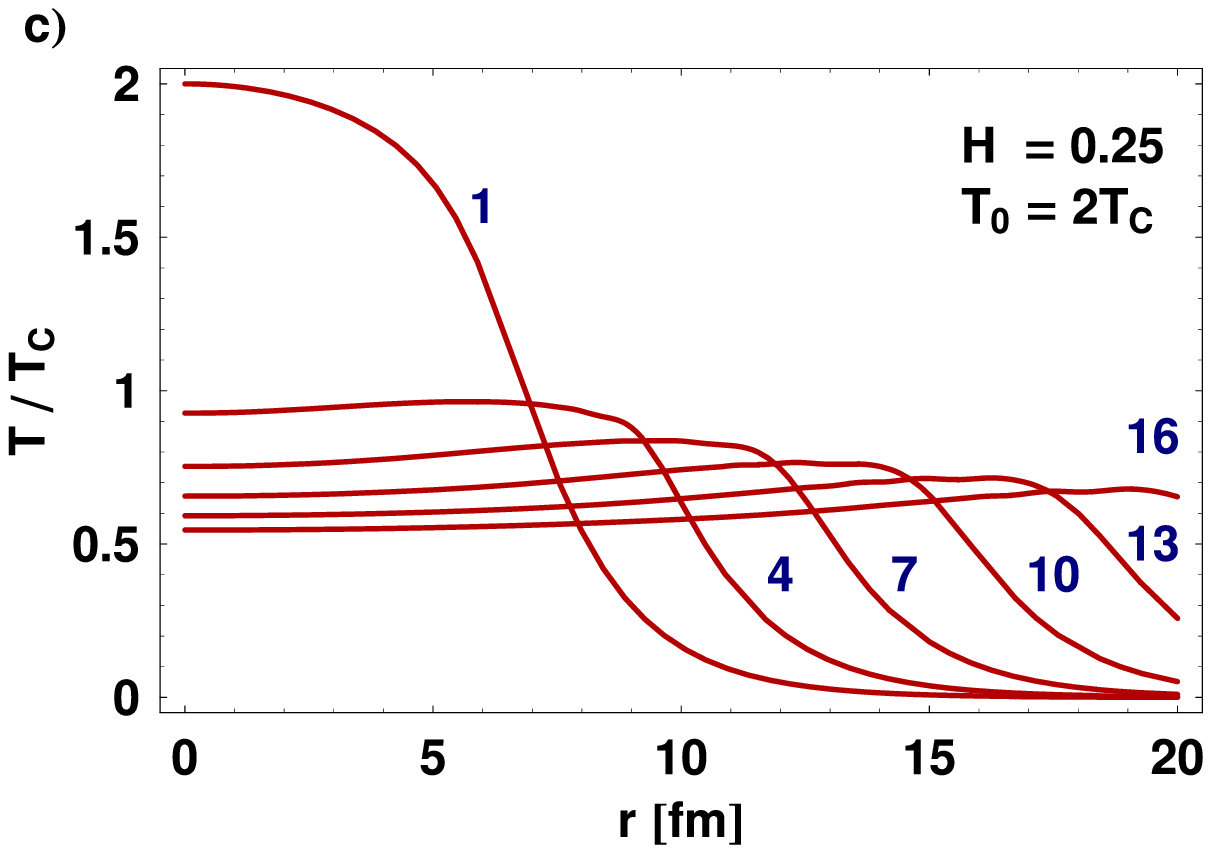}} 
\subfigure{\includegraphics[angle=0,width=0.4\textwidth]{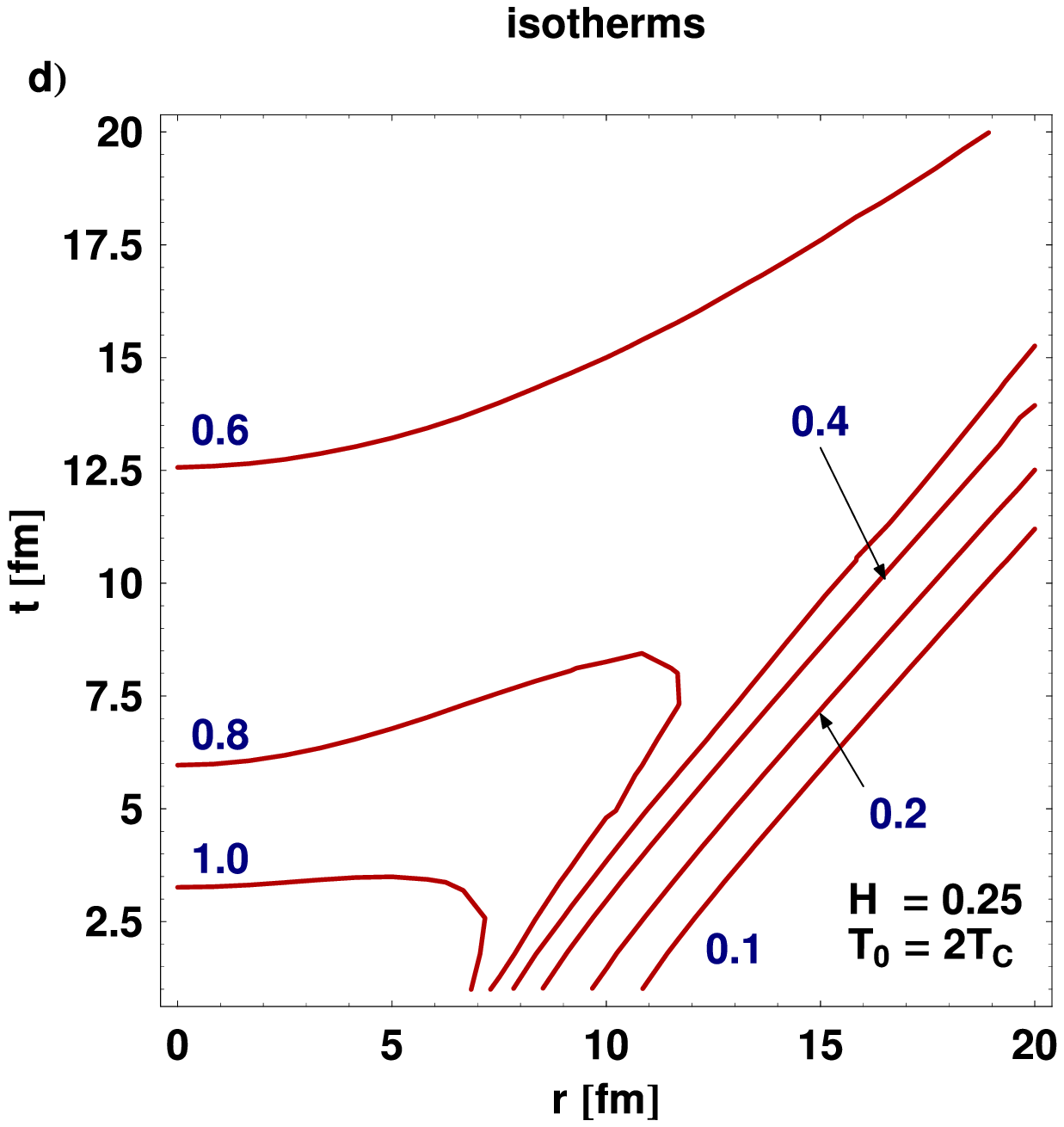}} \\
\end{center}
\caption{Hydrodynamic expansion of matter with initial pre-equilibrium
flow characterized by the velocity profile (\ref{initvr}) with
$H=0.25$. The initial central temperature $T_0 = 2 T_c$. 
Notation as in Fig. \ref{fig:HIN2TH00}.}
\label{fig:HIN2TH25}
\end{figure}

In Figs. \ref{fig:HIN2TH00} and \ref{fig:HIN2TH25} we show our 
results obtained for two different initial conditions characterized by the two different 
values of the parameter $H$, $H = 0$ and $H=0.25 \hbox{\,fm}^{-1}$, respectively. 
In both cases the lattice equation of state is used and the initial temperature 
in the center of the system is assumed to be twice the temperature of the phase 
transition, $T_0 = T(t=t_0, r=0) = 2 T_c$. This means, that for the commonly accepted 
value of  $T_c$, which is about 170 MeV, the initial temperature in the center 
reaches about 340 MeV.

In the case $H=0$, the transverse flow is initially set to zero but it builds up
during the evolution of the system, as shown in Fig. \ref{fig:HIN2TH00} a). 
The corresponding values of the transverse four-velocity $u_r$ are plotted in the part b).
To check if the flow approaches the asymptotic scaling solution, we compare the 
velocity profiles calculated numerically at different times $t_i = 1,4,7,10,13,16$ fm
(solid lines)  with the ideal Hubble-like velocities of the form $r/t_i$
(thin dashed lines) in the regions $r < t_i$. As expected, in the case $H=0$ the calculated 
profiles do not agree with the ideal curves in the considered evolution times. 

In the part c) of Fig. \ref{fig:HIN2TH00} we observe that the central part of the system 
cools down very fast from $T = 2 T_c$ down to the critical temperature $T=T_c$ and the 
subsequent cooling is very much slowed down. This behavior is caused by different values
of the sound velocity in the regions above and below $T_c$; larger values of $c_s$ in the 
plasma phase imply its faster cooling. We note that for the first order phase transition 
the speed of sound drops to zero at $T= T_c$ and the system keeps on expanding at a constant 
temperature of $T=T_c$. In the present case we deal with a sudden cross-over transition, 
hence the expansion of the volume is coupled to a small decrease of the temperature. 
From Fig. \ref{fig:HIN2TH00} d) showing the isotherms in the $t-r$ space, we may conclude 
that expansion of the system without noticeable cooling below $T_c$ takes more than 20 fm.

The evolution of matter shown in Fig. \ref{fig:HIN2TH00} may be confronted with
the situation where the non-zero pre-equlibrium flow is included in the initial condition.
In Fig. \ref{fig:HIN2TH25} we show our results obtained in the case $H=0.25 \hbox{\,fm}^{-1}$, 
with the same initial central temperature $T_0 = 2 T_c$. Since the transverse flow is 
present already at the beginning of the evolution, the expansion of the system is much faster 
than that discussed in the previous case of $H=0$. In Fig. \ref{fig:HIN2TH25} a) we 
show the velocity profiles in $r$, again for 6 different values of time. 
By comparing to the ideal curves of the form $r/t_i$ (thin dashed lines)
we observe that the flow approaches the asymptotic scaling solution after about 7 fm.

In Fig. \ref{fig:HIN2TH25}  c) we can see similar behavior to that presented in
Fig. \ref{fig:HIN2TH00}  c), namely,  the initial fast cooling of the hot center, which
is slowed down when the system approaches $T_c$. We observe, however, that in the case 
$H=0.25 \hbox{\,fm}^{-1} $ the slowing down of the cooling process is not as much effective 
as that observed in the case $H=0$. Due to the larger transverse flow, the energy from 
the interior is transported outside, the temperature in the center continues to drop down, 
and a very interesting situation happens: the parts of the system away from the center become
hotter than the parts in the center. This phenomenon is well seen in Fig.
\ref{fig:HIN2TH25}  d) where the isotherms of the system are shown.
We note that isotherms of similar shape are used in the Cracow model  
where they are defined by the condition of constant proper time $\tau$
\cite{Broniowski:2001we}. On the other hand, the Blast-Wave model assumes 
that freeze-out happens at a constant value of the ordinary time $t$ with a fixed
temperature $T$;
whereas the Buda-Lund model assumes that the temperature
profile may eventually decrease to 0 at large transverse extensions, hence
capturing the feature that the temperature vanishes for very large
transverse coordinates in all of the presented calculations.

\begin{figure}[t]
\begin{center}
\subfigure{\includegraphics[angle=0,width=0.3\textwidth]{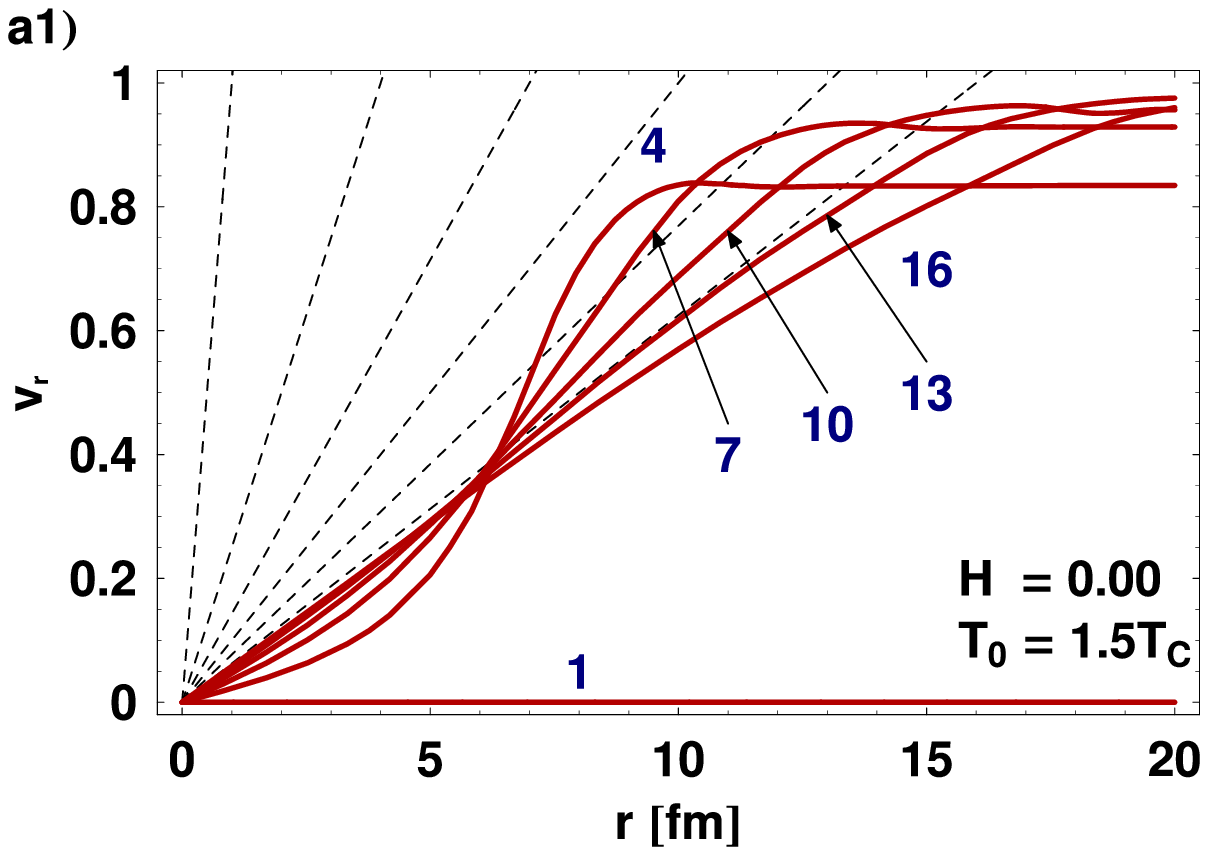}} 
\subfigure{\includegraphics[angle=0,width=0.3\textwidth]{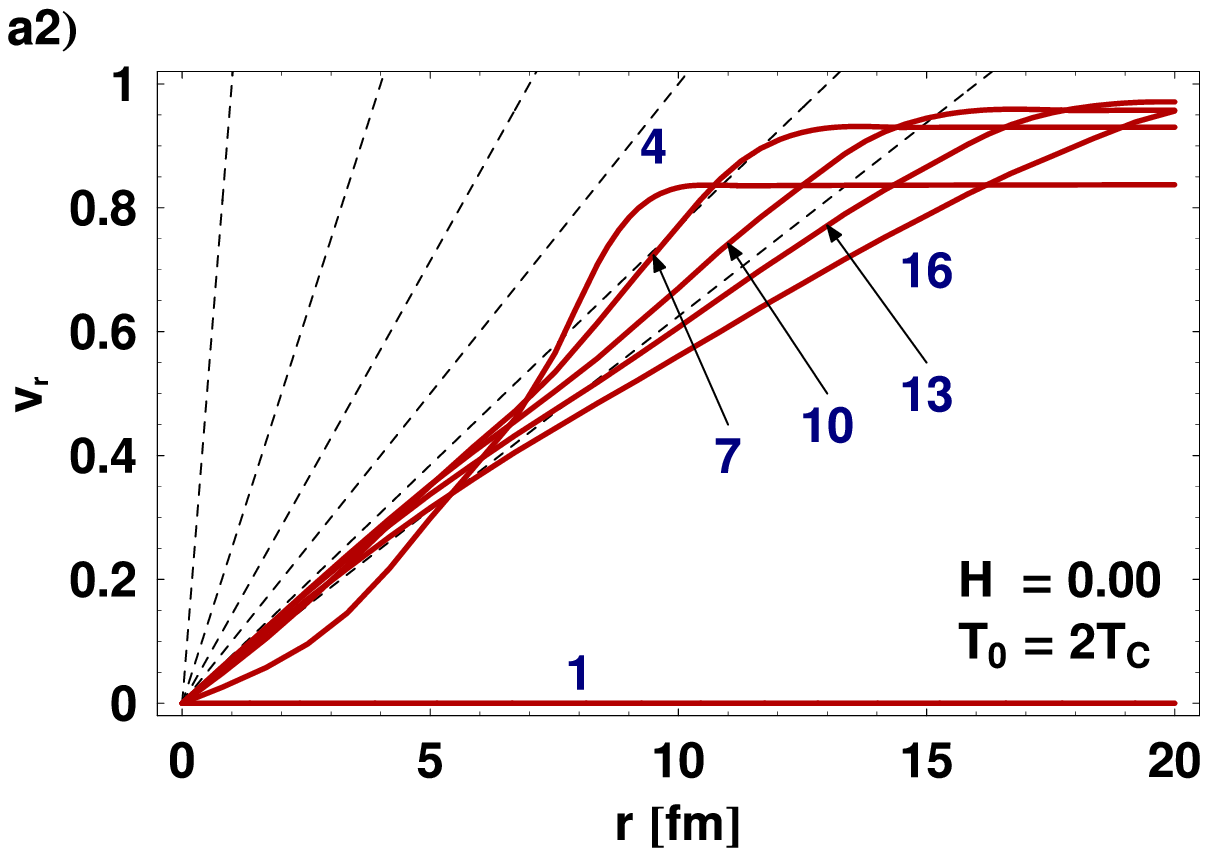}} 
\subfigure{\includegraphics[angle=0,width=0.3\textwidth]{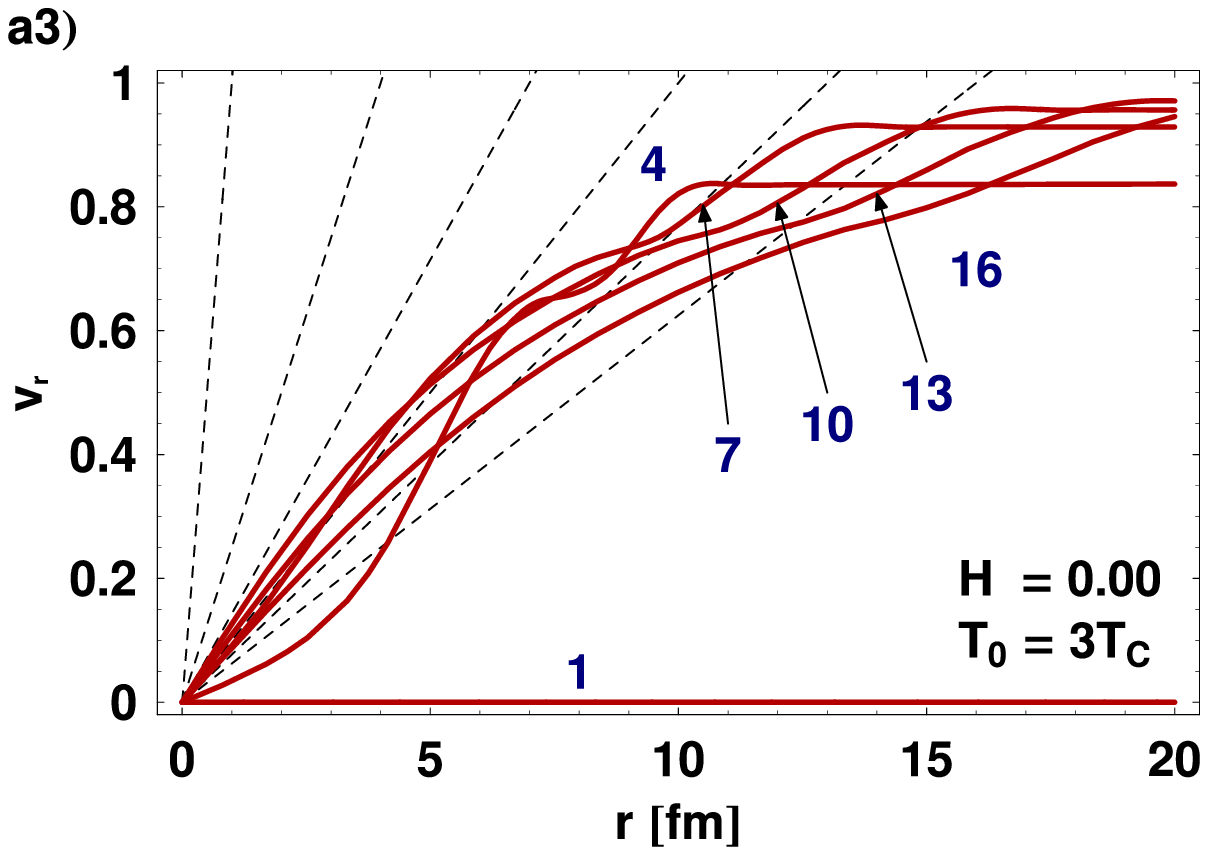}} \\
\subfigure{\includegraphics[angle=0,width=0.3\textwidth]{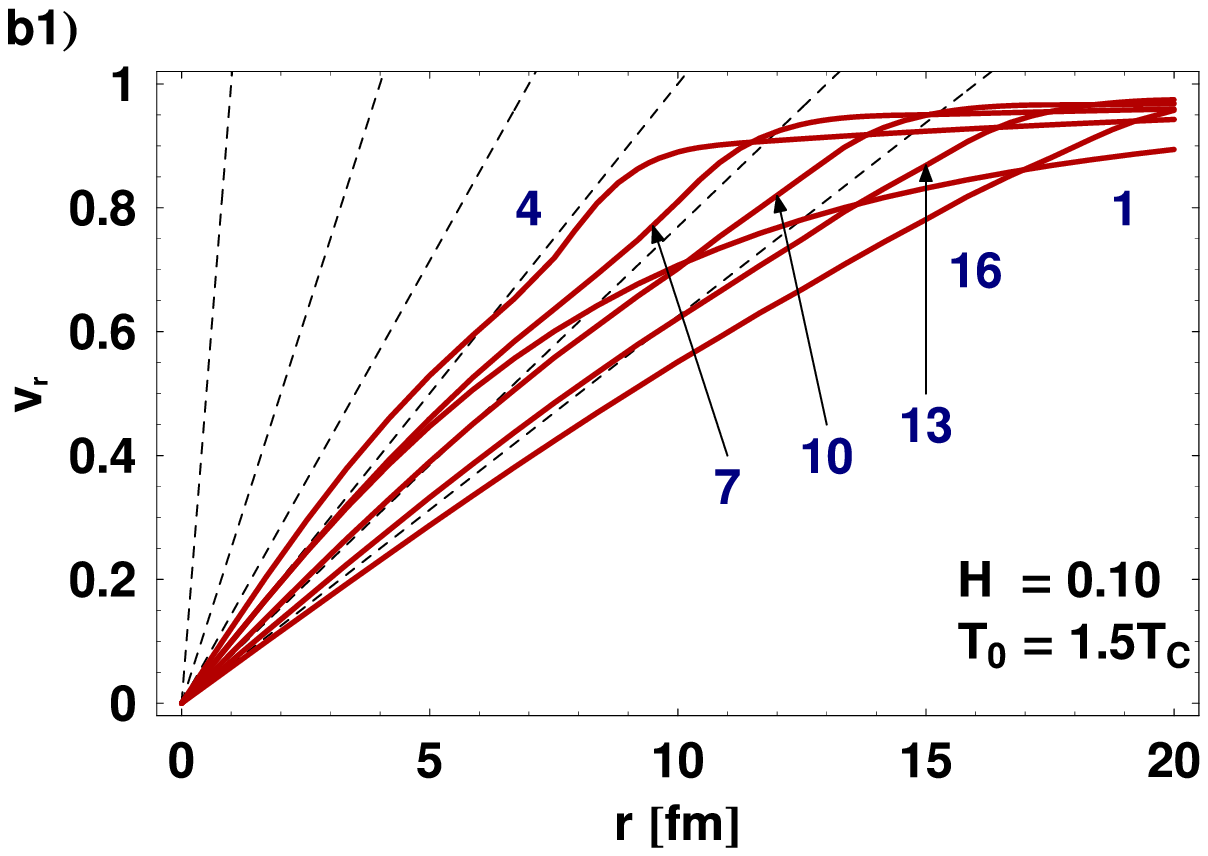}} 
\subfigure{\includegraphics[angle=0,width=0.3\textwidth]{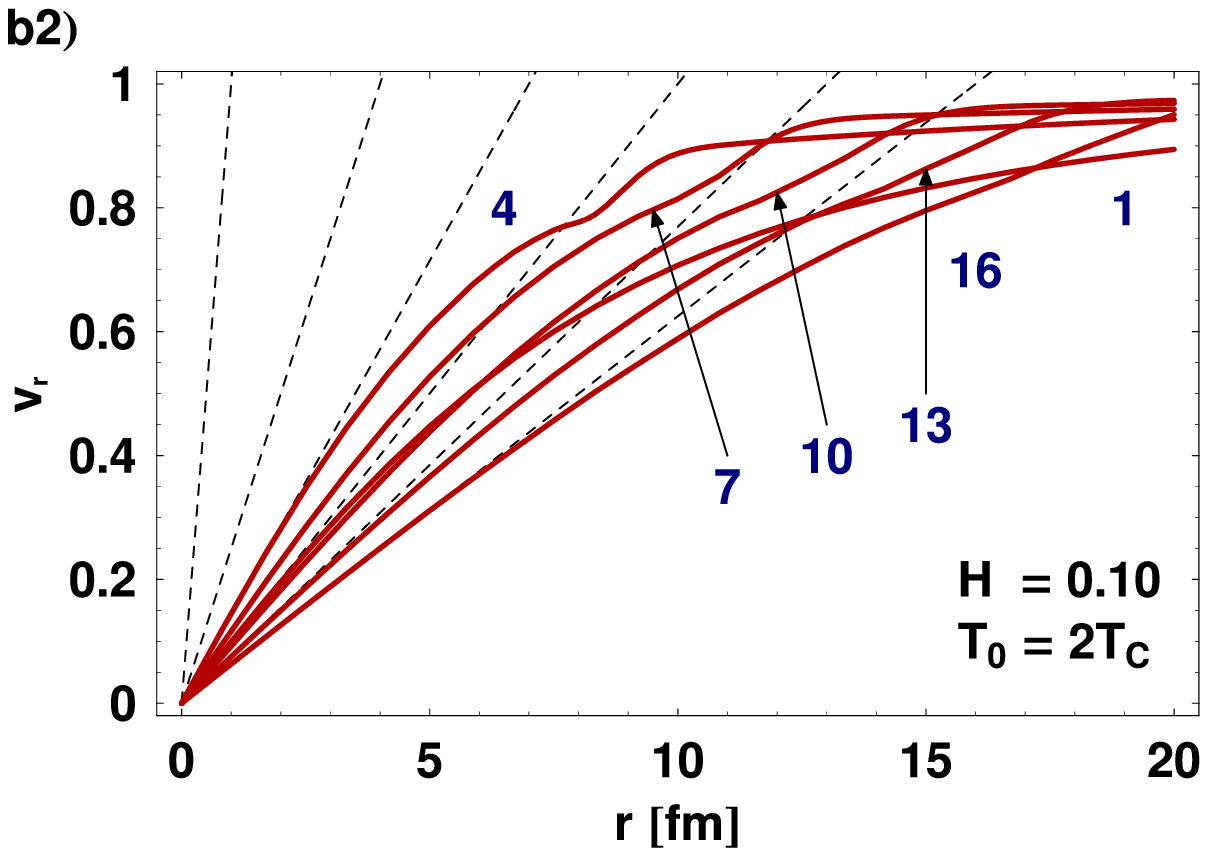}} 
\subfigure{\includegraphics[angle=0,width=0.3\textwidth]{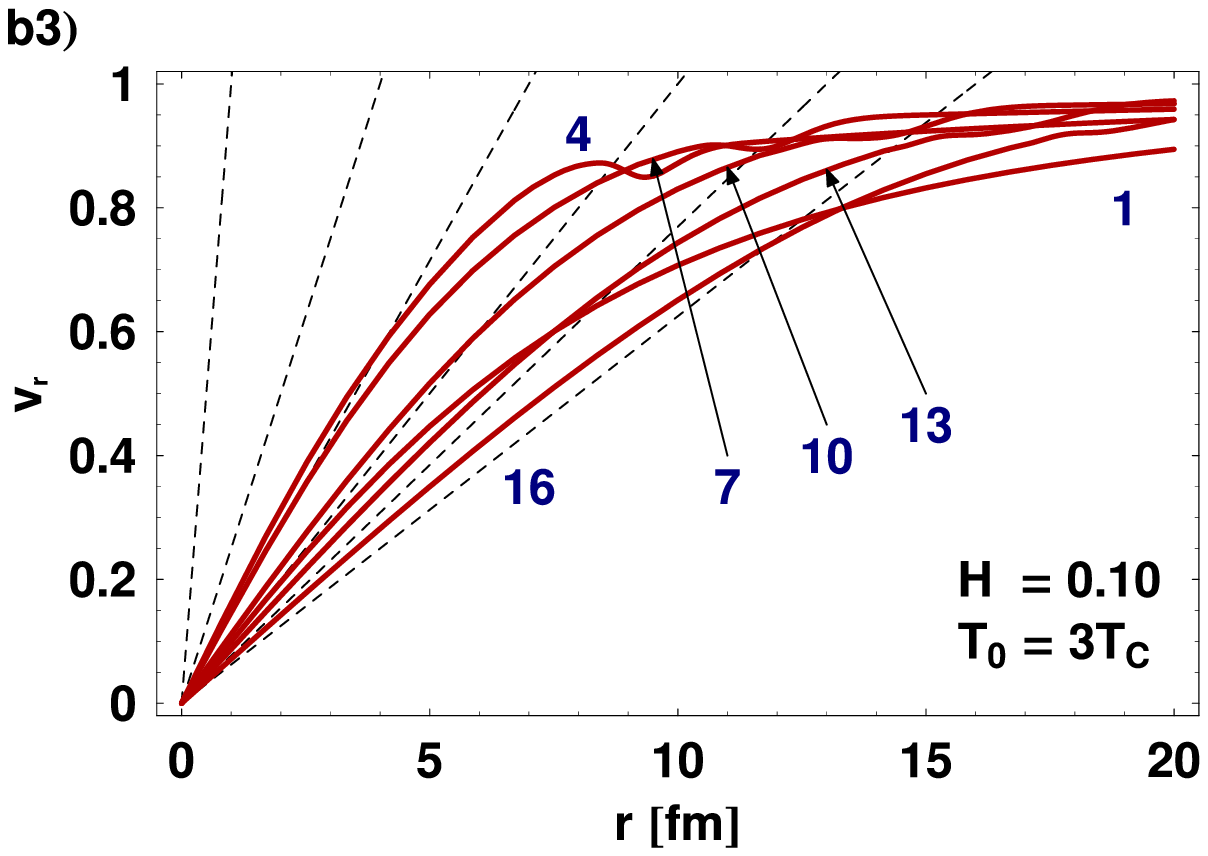}} \\
\subfigure{\includegraphics[angle=0,width=0.3\textwidth]{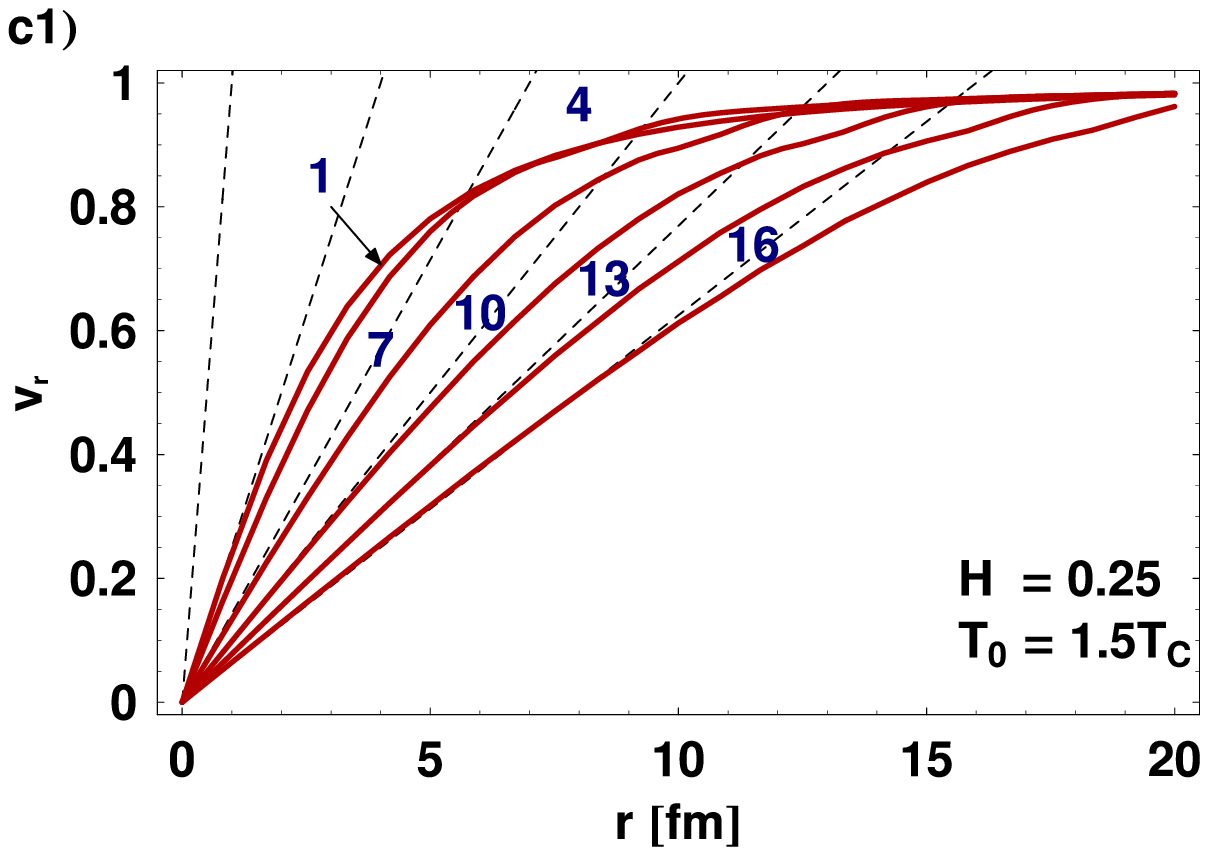}} 
\subfigure{\includegraphics[angle=0,width=0.3\textwidth]{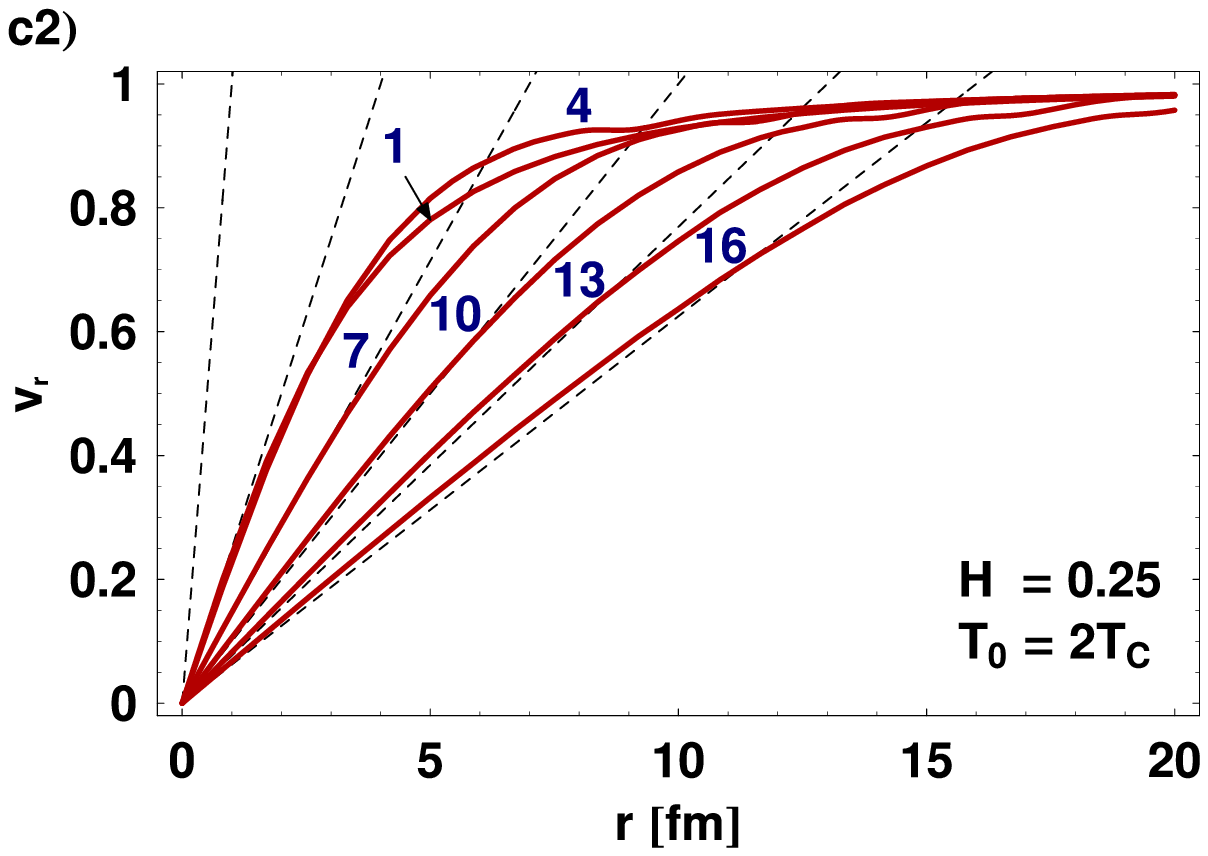}} 
\subfigure{\includegraphics[angle=0,width=0.3\textwidth]{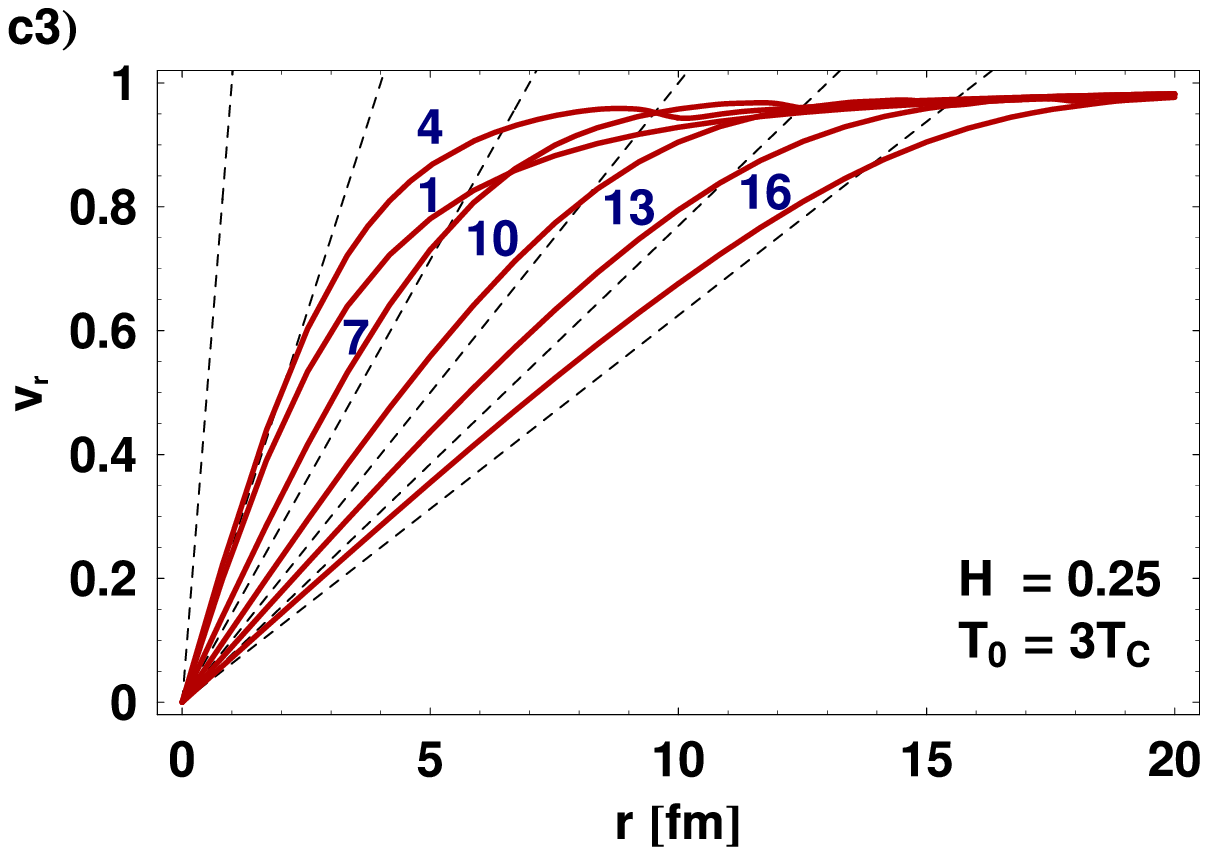}} \\
\subfigure{\includegraphics[angle=0,width=0.3\textwidth]{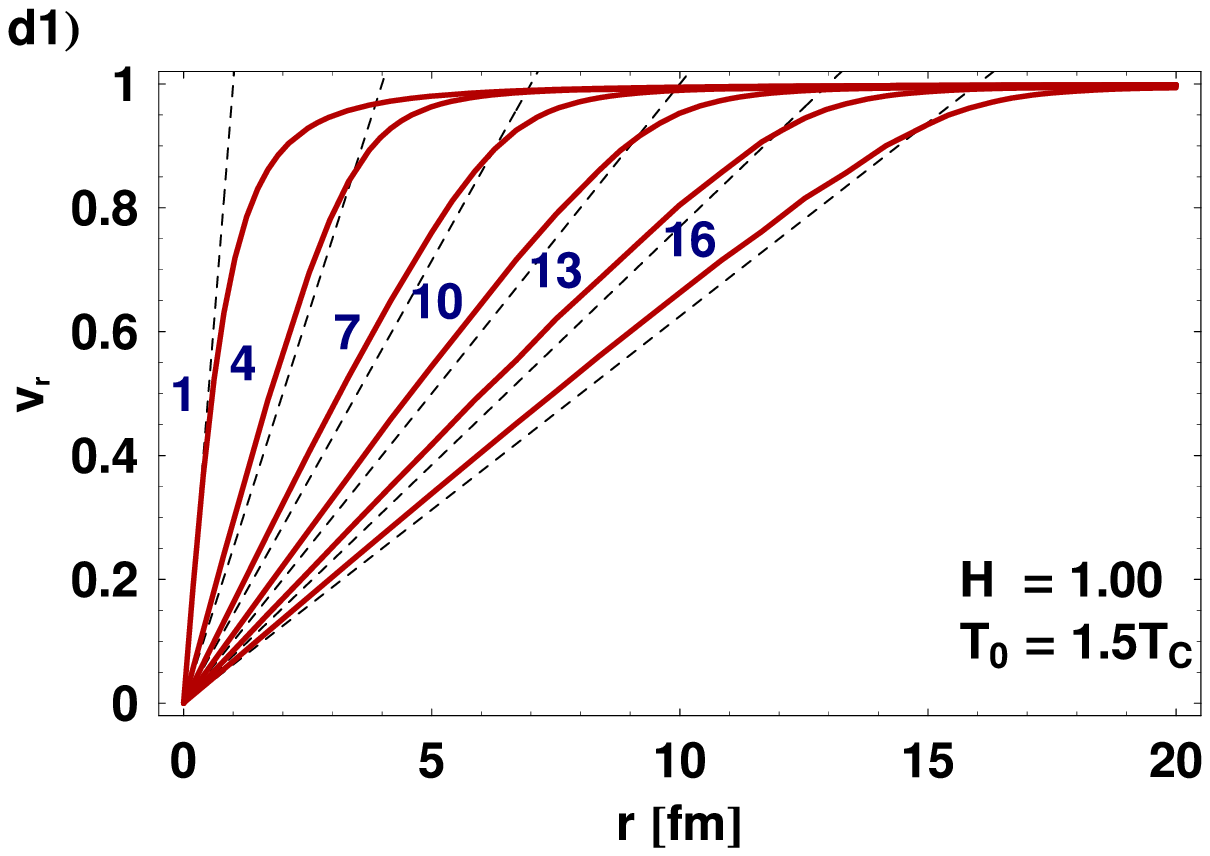}} 
\subfigure{\includegraphics[angle=0,width=0.3\textwidth]{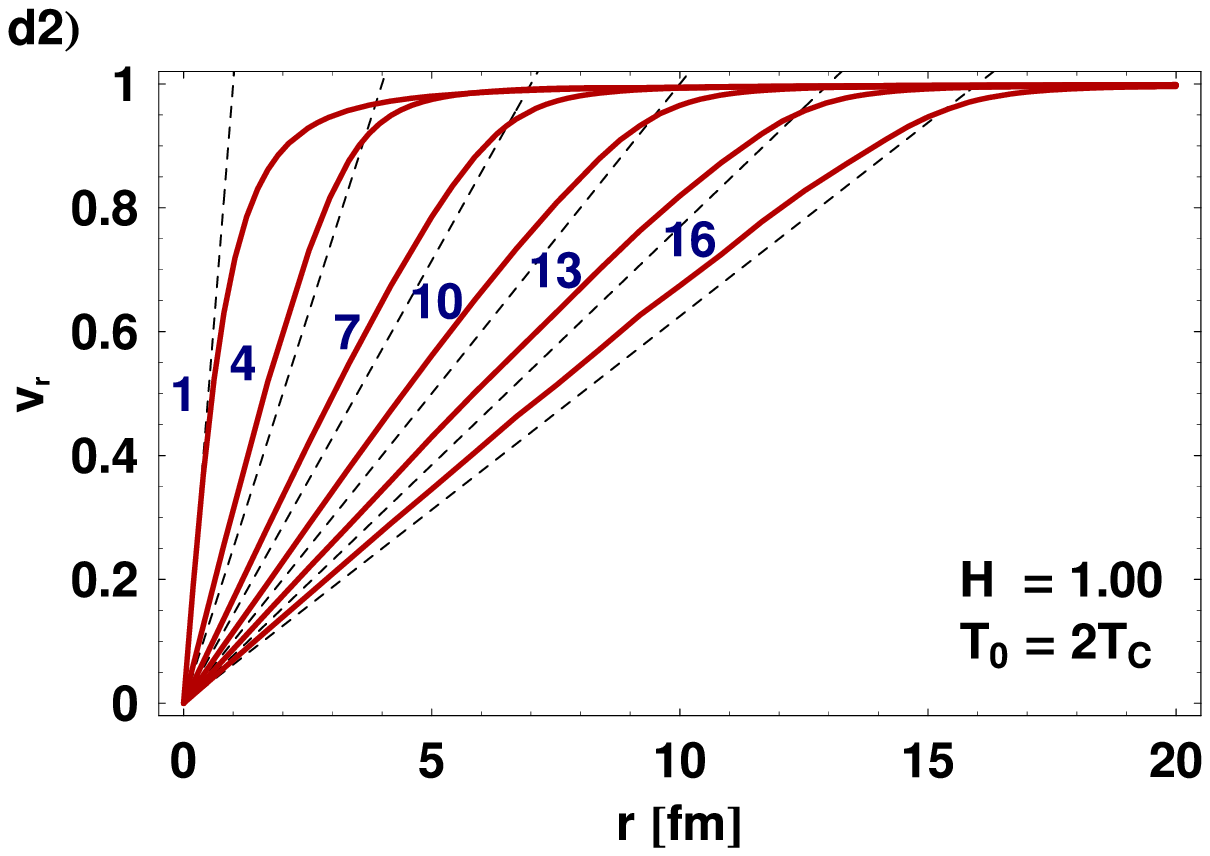}} 
\subfigure{\includegraphics[angle=0,width=0.3\textwidth]{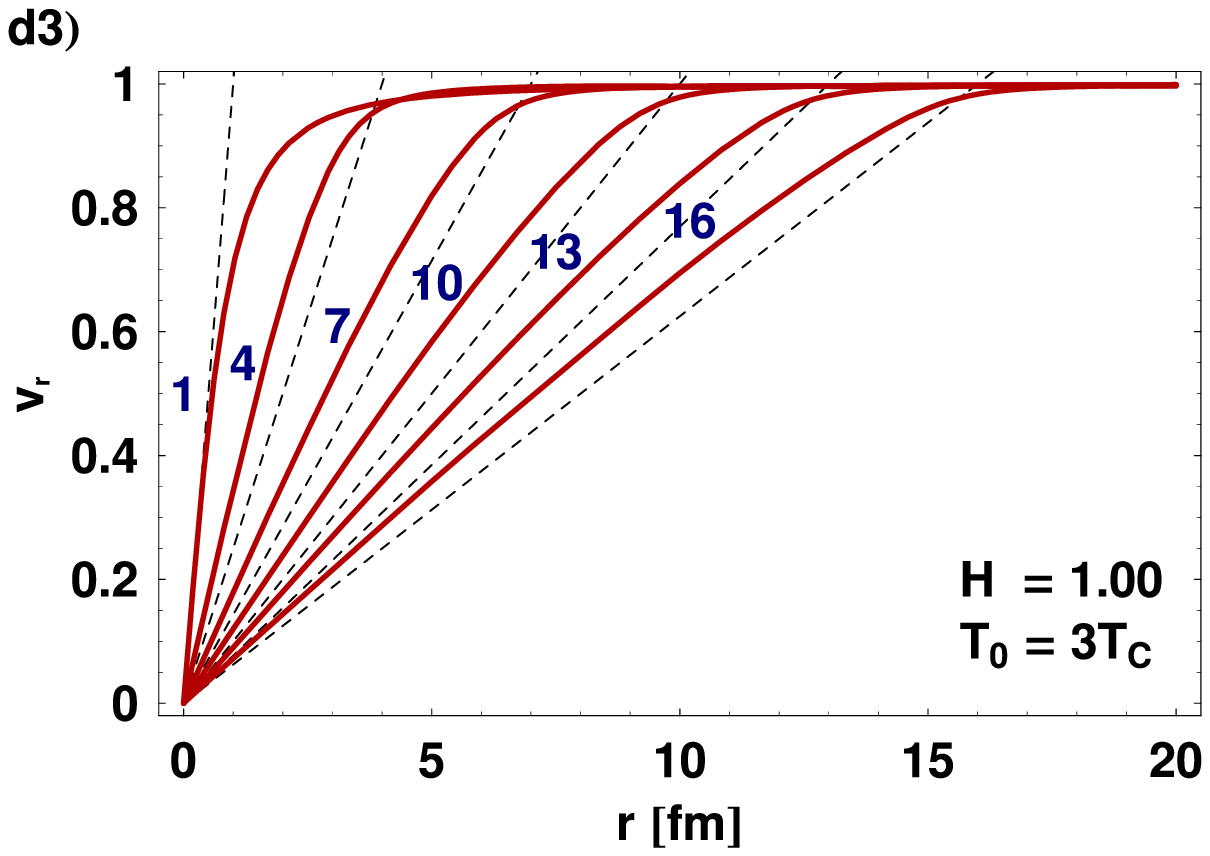}} \\
\end{center}
\caption{Velocity profiles in $r$ (solid lines) calculated for different values of the
parameter $H$ and different values of the initial central temperature $T_0$.
The labels 1,4,7,10,13 and 16 denote the evolution time in fm. The values of
$H$ are given in 1/fm. The thin dashed lines show the ideal profiles of the form $r/t$
obtained for the same values of time.}
\label{fig:flowprofiles}
\end{figure}

In Fig. \ref{fig:flowprofiles} we show a collection of the velocity profiles obtained
for four different values of the parameter $H$ and for three different values of the central
temperature $T_0$.  The four rows of smaller figures describe the results
obtained with $H = 0.0,\,0.10,\,0.25,\,1.0 \hbox{\,fm}^{-1}$,
while the three columns correspond to the central temperature $T_0 = 
1.5 T_c, \, T_0 = 2 T_c$, and $T_0 = 3 T_c$.

The results presented in Fig. \ref{fig:flowprofiles} are obtained with
the analytic model for the temperature dependence of the sound velocity, 
hence, by comparison of Fig. \ref{fig:flowprofiles} with the two previous figures,
the effects of the change of the function $c_s(T)$ on the
time evolution of the system may be observed. For example, comparing the part 
c2) of Fig.  \ref{fig:flowprofiles} with the part a) of Fig. \ref{fig:HIN2TH25} 
(both results obtained with $H = 0.25 \,\hbox{\,fm}^{-1}$ and $T_0 = 2 T_c$) we can see that 
the flow obtained with the analytic model is  closer to the asymptotic scaling form than the 
flow obtained for the lattice equation of state. We note that the main difference between 
the two cases is that the sound velocity in the analytic model is smaller in the 
hadronic phase. The asymptotic scaling solutions are more easily generated at the reduced values 
of the sound velocity as shown in Refs. \cite{Cooper:1975qi,Biro:1999eh,
Biro:2000nj}. This effect explains a better agreement of the generated transverse 
flow with the asymptotic scaling solution obtained in the analytic model.

Let us now discuss the impact of the initial temperature on the formation of the
transverse flow. Since the spatial extension of the system is the same in all
considered cases (roughly a diameter of the gold nucleus), a higher initial temperature 
implies a larger pressure gradient leading directly to the formation of the 
stronger flow. This effect is seen in Fig. \ref{fig:flowprofiles} if the results 
obtained with the same value of $H$ but different values of $T_0$
are compared. The presence of the pre-equilibrium transverse
flow also helps to develop the strong transverse flow, and this effect is
added to the effects of the pressure gradient. This is clearly seen in the case 
$H = 0.10 \,\hbox{\,fm}^{-1}$, depicted in the parts b1) - b3) of 
Fig. \ref{fig:flowprofiles}. At $t=16$ fm the flow in the range $r < 10$ fm 
is close but below the asymptotic scaling solution for $T_0 = 2 T_c$, and close but above 
the  asymptotic scaling  solution for $T_0 = 3 T_c$. 
For $T_0 = 1.5 T_c$ the flow is noticeably below the asymptotic scaling 
form.  We conclude that the asymptotic solution 
may be reached either from above or from below (in the region $r << t$),
depending on the value of the initial temperature. If the pre-equilibrium flow 
is strong  enough the asymptotic solution
is approached from above. This behavior is indicated by our results obtained
in the cases $H = 0.25 \,\hbox{\,fm}^{-1}$ and $H = 1.0 \,\hbox{\,fm}^{-1}$,
Fig. \ref{fig:flowprofiles} c1) - d3).
Interestingly, in the case $H = 1/t_0 = 1.0 \,\hbox{\,fm}^{-1} $, where
the initial flow agrees with the scaling form already at the beginning of the 
time evolution,
the pressure gradient accelerates the matter and the convergence to the
scaling form is delayed. On the other hand, the existence of a non-zero pre-equilibrium
flow seems to be a necessary condition for the formation of the accelerationless Hubble-like 
flows in
the evolution times of 10-15 fm. This is indicated by our results obtained with $H=0$
for different values of $T_0$, Fig. \ref{fig:flowprofiles} a1) - a3).
The amount of the pre-equilibrium flow required to
achieve the fast convergence to the asymptotic solution depends on the initial temperature
(pressure gradient); smaller values of $H$ may be compensated by larger values of $T_0$.
For example, at $t = 16$ fm the flow profiles are very similar in the cases: 
$H = 1 \,\hbox{\,fm}^{-1}$ and $T = 1.5 T_0$ (part d1), 
$H = 0.25 \,\hbox{\,fm}^{-1}$ and $T=2 T_c$ (part c2), 
and also $H = 0.1 \,\hbox{\,fm}^{-1}$ and $T = 3 T_c$ (part b3).
It is quite remarkable, that the presence of this pre-equilibrium flow
is required only to set the value of the Hubble constant to $H = 1/t$ i.e.
to reach the {\it asymptotic} scaling solution within a short time period,
    however, we find that  the linear flow profile, $v = H r$ develops
    with $H \neq 1/t$ in all the considered cases by about 7 fm/c,
    regardless of the details of the initial conditions.

\section{Conclusions}

Our results show a dynamical development of the scaling solutions
in the relativistic hydrodynamics applied to relativistic heavy-ion
collisions. For the realistic initial conditions connecting the amount of 
the initially produced entropy with the number of participating nucleons,
we find that a Hubble type linear transverse flow, characterized by
$v = H r$ develops by about 7 fm/c regardless of the details of the
initial conditions when varied numerically in a reasonable range.
However, ²it is more difficult for the transverse flow to achieve
the {\it asymptotic}, accelerationless scaling form:
 to reach the values of $H = 1/t$ within the evolution time of about 10 -15 fm.
  For this, the necessary conditions are thus stronger:
pre-equilibrium transverse flow has to be present already at the beginning of the 
hydrodynamic evolution at the initial time of 1 fm. The amount of the 
pre-equilibrium flow, required for the fast approach to the asymptotic solutions,
depends on the initial pressure gradients; for larger gradients smaller initial
flows are necessary.

The results of these calculations give support for using the phenomenological 
parameterizations of the freeze-out process such as the Blast-Wave, Buda-Lund, 
and Cracow models. Certainly, more work should be done to combine the output 
of the hydrodynamic calculations with the description of freeze-out used in
these models, however, we showed that such features as the linear flow profiles or
the isotherms defined by the constant value of the proper time appear as the solutions 
of the relativistic hydrodynamics with suitably chosen initial conditions. 

\begin{acknowledgments}
This work was supported in part by the Polish State Committee of Scientific Research, 
grant 2~P03B~05925, by the Hungarian OTKA grant T038406, the MTA - OTKA - NSF
grant INT0089462 and 
by the exchange program of the Hungarian and Polish Academy of Sciences as well as 
by the Polish KBN - Hungarian Ministry of Education Exchange Programme in Science and Technology.
\end{acknowledgments}


\begin{thebibliography}{10}
\expandafter\ifx\csname url\endcsname\relax
  \def\url#1{\texttt{#1}}\fi
\expandafter\ifx\csname urlprefix\endcsname\relax\def\urlprefix{URL }\fi

\bibitem{QM02}
QM02, Nucl. Phys., {\bf A715}, ~Proc. of the 16th Int. Conf. on
  Ultra-Relativistic Nucleus-Nucleus Collisions, Nantes, France.

\bibitem{Broniowski:2001we}
W.~Broniowski, W.~Florkowski, Phys. Rev. Lett., {\bf 87} (2001) 272302,
  nucl-th/0106050.

\bibitem{Broniowski:2001uk}
W.~Broniowski, W.~Florkowski, Phys. Rev., {\bf C65} (2002) 064905,
  nucl-th/0112043.

\bibitem{Csanad:2004mm}
M.~Csan\'ad, T.~Cs\"org\H{o}, B.~L\"orstad, A.~Ster, J. Phys., {\bf G30} (2004)
  S1079--S1082, nucl-th/0403074.

\bibitem{allday}
	J. Allday, ``Quarks, Leptons and the Big Bang",
	IOP Publishing Ltd 2002, ISBN 0 7503 0806 0

\bibitem{zbg}
	J. P. Bondorf, S. I. A. Garpman and J. Zim\'anyi,
	Nucl. Phys. {\bf A296} (1978) 320-332

\bibitem{Landau:1953gs}
L.~D. Landau, Izv. Akad. Nauk SSSR Ser. Fiz., {\bf 17} (1953) 51--64.

\bibitem{Belenkij:1956cd}
S.~Z. Belenkij, L.~D. Landau, Nuovo Cim. Suppl., {\bf 3S10} (1956) 15.

\bibitem{Cooper:1974ak}
F.~Cooper, G.~Frye, E.~Schonberg, Phys. Rev. Lett., {\bf 32} (1974) 862.

\bibitem{Cooper:1975qi}
F.~Cooper, G.~Frye, E.~Schonberg, Phys. Rev., {\bf D11} (1975) 192.

\bibitem{Bjorken:1983qr}
J.~D. Bjorken, Phys. Rev., {\bf D27} (1983) 140--151.

\bibitem{Baym:1983sr}
G.~Baym, B.~L. Friman, J.~P. Blaizot, M.~Soyeur, W.~Czyz, Nucl. Phys., {\bf
  A407} (1983) 541--570.

\bibitem{Hwa:1974gn}
R.~C. Hwa, Phys. Rev., {\bf D10} (1974) 2260.

\bibitem{Chiu:1975hx}
C.~B. Chiu, K.-H. Wang, Phys. Rev., {\bf D12} (1975) 272.

\bibitem{Chiu:1975hw}
C.~B. Chiu, E.~C.~G. Sudarshan, K.-H. Wang, Phys. Rev., {\bf D12} (1975) 902.

\bibitem{Gorenstein:1977xv}
M.~I. Gorenstein, Y.~M. Sinyukov, V.~I. Zhdanov, Phys. Lett., {\bf B71} (1977)
  199--202.

\bibitem{Gorenstein:1978jg}
M.~I. Gorenstein, Y.~M. Sinyukov, V.~I. Zhdanov, Zh. Eksp. Teor. Fiz., {\bf 74}
  (1978) 833--845.

\bibitem{Huovinen:2002fp}
P.~Huovinen, Acta Phys. Polon., {\bf B33} (2002) 1635--1650, nucl-th/0204029.

\bibitem{Kolb:2003dz}
P.~F. Kolb, U.~Heinz, nucl-th/0305084.

\bibitem{Teaney:2001av}
D.~Teaney, J.~Lauret, E.~V. Shuryak, nucl-th/0110037.

\bibitem{Hirano:2004er}
T.~Hirano, J. Phys., {\bf G30} (2004) S845--S852, nucl-th/0403042.

\bibitem{Biro:1999eh}
T.~S. Biro, Phys. Lett., {\bf B474} (2000) 21--26, nucl-th/9911004.

\bibitem{Biro:2000nj}
T.~S. Biro, Phys. Lett., {\bf B487} (2000) 133--139, nucl-th/0003027.

\bibitem{Csorgo:2002ki}
T.~Cs\"org\H{o}, F.~Grassi, Y.~Hama, T.~Kodama, Heavy Ion Physics, {\bf A21} (2004)
  53--62, hep-ph/0203204.

\bibitem{Csorgo:2002bi}
T.~Cs\"org\H{o}, F.~Grassi, Y.~Hama, T.~Kodama, Heavy Ion Physics, {\bf A21} (2004)
  63--71, hep-ph/0204300.

\bibitem{Csorgo:2003rt}
T.~Cs\"org\H{o}, F.~Grassi, Y.~Hama, T.~Kodama, Phys. Lett., {\bf B565} (2003)
  107--115, nucl-th/0305059.

\bibitem{Csorgo:2003ry}
T.~Cs\"org\H{o}, L.~P. Csernai, Y.~Hama, T.~Kodama, Heavy Ion Physics, {\bf A21}
  (2004) 73--84, nucl-th/0306004.

\bibitem{Retiere:2003kf}
F.~Retiere, M.~A. Lisa, nucl-th/0312024.

\bibitem{Retiere:2004wa}
F.~Retiere, J. Phys., {\bf G30} (2004) S827--S834, nucl-ex/0405024.

\bibitem{Makhlin:1987gm}
A.~N. Makhlin, Y.~M. Sinyukov, Z. Phys., {\bf C39} (1988) 69.

\bibitem{Kolb:2002ve}
P.~F. Kolb, R.~Rapp, Phys. Rev., {\bf C67} (2003) 044903, hep-ph/0210222.

\bibitem{Akkelin:2000ex}
S.~V. Akkelin, T.~Cs\"org\H{o}, B.~Luk\'acs, Y.~M. Sinyukov, M.~Weiner, Phys. Lett.,
  {\bf B505} (2001) 64--70, hep-ph/0012127.

\bibitem{Csorgo:2001ru}
T.~Cs\"org\H{o}, Central European Journal of Physics, {\bf 2 (4)} (2004) 1--10,
  hep-ph/0111139.

\bibitem{Csorgo:2001xm}
T.~Cs\"org\H{o}, S.~V. Akkelin, Y.~Hama, B.~Luk\'acs, Y.~M. Sinyukov, Phys. Rev., {\bf
  C67} (2003) 034904, hep-ph/0108067.

\bibitem{Bearden:2003fw}
I.~G. Bearden, {\it et~al.}, BRAHMS, Phys. Rev. Lett., {\bf 90} (2003) 102301.

\bibitem{Bearden:2004yx}
I.~G. Bearden, BRAHMS, nucl-ex/0403050.

\bibitem{Florkowski:2001fp}
W.~Florkowski, W.~Broniowski, M.~Michalec, Acta Phys. Polon., {\bf B33} (2002)
  761--769, nucl-th/0106009.

\bibitem{Torrieri:2004zz}
G.~Torrieri, W.~Broniowski, W.~Florkowski, J.~Letessier, J.~Rafelski,
  nucl-th/0404083.

\bibitem{Mohanty:2003va}
B.~Mohanty, J.-e. Alam, Phys. Rev., {\bf C68} (2003) 064903, nucl-th/0301086.

\bibitem{Karsch:2001vs}
F.~Karsch, Nucl. Phys., {\bf A698} (2002) 199--208, hep-ph/0103314.

\bibitem{Szabo:2003kg}
K.~K. Szab\'o, A.~I. T\'oth, JHEP, {\bf 06} (2003) 008, hep-ph/0302255.

\bibitem{Broniowski:2000bj}
W.~Broniowski, W.~Florkowski, Phys. Lett., {\bf B490} (2000) 223--227,
  hep-ph/0004104.

\bibitem{Broniowski:2000hd}
W.~Broniowski, hep-ph/0008112.

\bibitem{Broniowski:2004yh}
W.~Broniowski, W.~Florkowski, L.~Y. Glozman, hep-ph/0407290.

\bibitem{Hagedorn:1965st}
R.~Hagedorn, Nuovo Cim. Suppl., {\bf 3} (1965) 147--186.

\bibitem{Prorok:2002ht}
D.~Prorok, Acta Phys. Polon., {\bf B33} (2002) 1583--1600, hep-ph/0205221.

\end{thebibliography}
\end{document}